\documentclass[aps,amsmath,amssymb,prd,showpacs,floatfix,preprint,superscriptaddress,nofootinbib,12pt]{article}
\pdfoutput=1
\usepackage{jheppub}
\usepackage{slashed}
\usepackage{comment}
\usepackage{fancyhdr}
\makeatletter
\def\@fpheader{\relax}
\makeatother
\usepackage[svgnames]{xcolor}
\usepackage[framemethod=tikz]{mdframed}
\usepackage{amsmath,epsfig}
\usepackage{amssymb,amsfonts}
\usepackage{latexsym}
\usepackage{subeqnarray}
\usepackage{xcolor}

\usepackage{graphicx}
\usepackage{ulem}
\usepackage{bbold}
\usepackage[utf8]{inputenc}

\newcommand{\cool}{\ensuremath{%
  \mathchoice{\includegraphics[height=3ex]{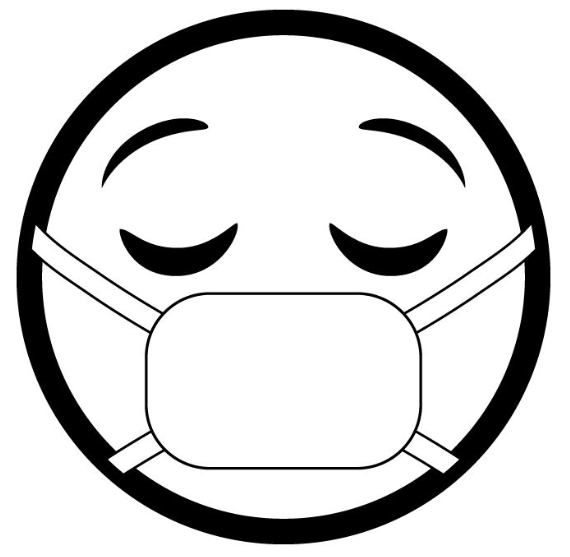}}
    {\includegraphics[height=3ex]{cattura.png}}
    {\includegraphics[height=2.5ex]{cattura.png}}
    {\includegraphics[height=2ex]{cattura.png}}
}}

\relax
\def\be{\begin{equation}}
\def\ee{\end{equation}}
\def\bea{\begin{eqnarray}}
\def\eea{\end{eqnarray}}
\newcommand\fverb{\setbox\pippobox=\hbox\bgroup\verb}
\newcommand\fverbdo{\egroup\medskip\noindent%
                        \fbox{\unhbox\pippobox}\ }
\newcommand\fverbit{\egroup\item[\fbox{\unhbox\pippobox}]}

\newcommand{\bear}{\begin{eqnarray}}

\newcommand{\eear}{\end{eqnarray}}

\newcommand{\bsea}{\begin{subeqnarray}}
\newcommand{\esea}{\end{subeqnarray}}
\newbox\pippobox

\def\6{\partial}

\newcommand{\comments}[1]{}
\newcommand{\blue}[1]{{\color{blue} #1 \color{black}}}

\input Starburst.fd

\newcommand{\OP}[1]{{\blue{OP: #1}}} % Oriol

\allowdisplaybreaks[3]

\preprint{IFT-UAM/CSIC-19-37}

\begin{document}

\title{\centering \LARGE Black Rubber and the
Non-linear\\ Elastic Response of Scale Invariant Solids}% Force line breaks with \\

\author[\,\,\cool]{Matteo Baggioli}

\affiliation[\cool]{Instituto de Fisica Teorica UAM/CSIC,
c/ Nicolas Cabrera 13-15, Cantoblanco, 28049 Madrid, Spain}

\author[\,\,\star]{, Víctor Cáncer Castillo}
\affiliation[\star]{Institut de F\'isica d'Altes Energies (IFAE), The Barcelona Institute of Science and Technology (BIST), 
Campus UAB, 08193 Bellaterra, Barcelona.}

\author[\,\,\star]{, Oriol Pujol{\`a}s}

\emailAdd{matteobaggioli@uam.es}
\emailAdd{vcancer@ifae.es}
\emailAdd{pujolas@ifae.es}

\vspace{1cm}

\abstract{
We discuss the nonlinear elastic response in scale invariant solids. 
Following previous work, we split the analysis into two basic options: according to whether scale invariance (SI) is a manifest or a spontaneously broken symmetry.
In the latter case, one can employ effective field theory methods, whereas in the former we use holographic methods. We focus on a simple class of holographic models that exhibit elastic behaviour, and obtain their nonlinear stress-strain curves as well as an estimate of the elasticity bounds — the maximum possible deformation in the elastic (reversible) regime. 
The bounds differ substantially in the manifest or spontaneously broken SI cases, even when the same stress-strain curve is assumed in both cases. 
Additionally, the hyper-elastic subset of models (that allow for large deformations) is found to have stress-strain curves akin to natural rubber.
The holographic instances in this category, which we dub {\it black rubber}, display richer stress-strain curves -- with two different power-law regimes at different magnitudes of the strain. 
}

\maketitle

\section{Introduction}

The response of materials under mechanical (or elastic) deformations is a basic aspect of matter, which is important to understand and characterize. This is an old field of study because of the many practical applications and a large part of it is well understood since long ago \cite{landau7,Lubensky}. 

\begin{figure}[t]
\begin{center}
\includegraphics[width=9cm]{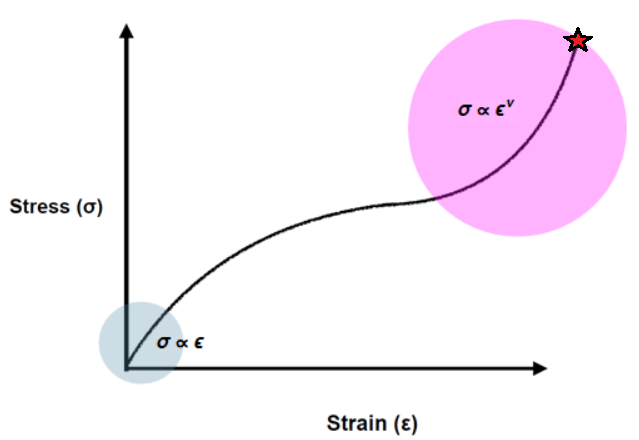}
\caption{Cartoon of a typical stress-strain curve of a hyper-elastic material \cite{enlighten70333}. In light blue shade the linear regime, in which stress $\propto$ strain.
At large strain deformation, the stress-strain relation can display a power law behaviour, $\sigma \propto \epsilon^\nu$, with some exponent $\nu$. The light red area illustrates this behaviour with $\nu>1$. Materials typically break or fracture after some critical deformation, which translates in the stress-train curves terminating at some point. The red star indicates the breaking point.
}
 \label{stressstrain}
\end{center}
\end{figure}

The response is best understood when restricted to the `linear' regime (small deformations), but there are many examples of solids that can undertake large deformations \cite{Ogden2004}. Common examples of these are the rubbers, but more generically they are referred to as hyper-elastic materials. The non-linear response that these materials exhibit is encoded in the {\it stress-strain relations} --  the amount of constant stress that must be applied in order to deform by a certain amount the material --  see Fig.~\ref{stressstrain} for a prototypical example. 
These curves can be easily obtained from experiments, but they are usually difficult to compute from the microscopic ingredients (even within the reversible regime, that is neglecting plasticity and dissipative effects),  especially so in strongly coupled materials.
Moreover, the nonlinear response is characterized by a rather large number of parameters/observables ({\it e.g.}, all the derivatives of the stress-strain curve at the origin, the maximum strain that the material can undertake, etc), and it might well be that there exist correlations between them. 
This motivates the study of the nonlinear mechanical response using effective low energy methods, which on their own might capture some of these correlations and even how these parameters depend on external parameters (temperature, etc).

The `continuum limit' description of mechanical deformations represents one such effective method that is useful for nonlinear  response. This approach embodies already a broad literature, see \cite{ZAMM:ZAMM19850650903} for a review. As in hydrodynamics, the medium is described by a 3-vector  $\pi_i(t,x^j)$, the displacement vector of the solid {\it elements}. 
How the material  deforms is encoded in its gradient, the so-called strain tensor $\epsilon_{ij}\sim\partial_{(i} \pi_{j)}$.
The main difference between solids and fluids in this language is that a solid responds to a constant  external stress $\sigma_{ij}$ with constant $\epsilon_{ij}$, whereas a viscous fluid responds with a constant strain rate, $\dot \epsilon_{ij}$. The punchline, though, is that the same kind of effective description is possible both for fluids and solids at small frequencies (and momenta) \cite{PhysRevA.6.2401}.

For small applied stresses the response is linear, $\sigma_{ij}\propto\epsilon_{ij}$, and the proportionality constants are usually called  elastic moduli.
Nonlinear elasticity concerns the relation between the stress and strain tensors beyond the linear approximation.
The mathematical formalism required for this in the continuum limit was developed long ago, see \cite{ZAMM:ZAMM19850650903} for a comprehensive review. %
This results especially relevant for hyper-elastic materials or elastomers, which allow for large reversible deformations.
For them, the continuum limit description takes the form of a nonlinear theory for the displacement vector field $\pi_i$ that can be specified by an energy function (how the energy density depends on $\epsilon_{ij}$) or a (nonlinear) constitutive relation.

Symmetries allow to  translate nonlinear elasticity into quantum field theory language. Given that condensed matter breaks spontaneously spacetime translations and boosts, it is possible to catch the dynamics for the lightest degrees of freedom using the methods of Effective Field Theory (EFT) applied to the spontaneous breaking of these symmetries.  In solids, the Goldstone bosons associated to this spontaneous breaking can be identified as the (`acoustic') phonons   \cite{Leutwyler:1996er,Dubovsky:2011sj}. These phonons are the fluctuations in the displacement vector $\pi_i$, they are indeed gapless and therefore belong to the lowest lying excitations, which makes the whole EFT construction consistent. 

The possible form of the full nonlinear effective Lagrangian for the phonon fields can be obtained using the coset construction applied to spacetime symmetries, see \cite{Nicolis:2013lma,Nicolis:2015sra}. 
It was recognized in \cite{Alberte:2018doe} that, for solids, the resulting phonon effective Lagrangian takes precisely the same form as the continuum limit nonlinear elasticity theory for the displacement vector $\pi^i(t,x^j)$ for hyper-elastic materials \cite{ZAMM:ZAMM19850650903} (at leading order in derivatives). The crucial advantage is that the {\it solid EFTs} are full-fledged effective Lagrangians that include dynamical effects and relativistic corrections among other improvements \cite{Alberte:2018doe}.

Let us emphasize that, even if they are not formulated in terms of the microscopic degrees of freedom, the EFT methods have stringent predictive and constraining power. 
This point was illustrated in \cite{Alberte:2018doe}, where the correlation between various nonlinear observables was made manifest in the form of {\it elasticity bounds} -- limits on the strain that a material of certain type can possibly withstand depending on other properties of the material. The simplest example arises by considering a class of materials characterized by power-law scaling in the stress-strain curve, schematically, 
\begin{equation}
    \sigma \sim \epsilon^\nu\,,
\end{equation}
with some arbitrary exponent $\nu$. Some elastomers in nature follow such a power-law at large deformation \cite{ZAMM:ZAMM19850650903} with a variety of exponents. 
For a general  analysis of the elastic response, one can treat $\nu$ simply as an effective parameter to describe the nonlinear response (at least in a  class of materials). Interestingly, assuming this power-law response is enough to place {\it a priori} an upper limit on the maximum strain deformation, $\epsilon_{max}$, that the material can undertake \cite{Alberte:2018doe}. This maximum deformation $\epsilon_{max}$ plays the role of (an upper limit to) the mechanical breaking or failure point of materials. 

A nontrivial outcome of the EFT methods is that one can establish a relation between these two nonlinear elasticity parameter,  $\nu$ and $\epsilon_{max}$.
Let us emphasize two points in order to highlight the potential value of the EFT methodology for nonlinear elasticity.
First, we stress that the correlation between $\nu$ and $\epsilon_{max}$ is entirely based on low energy EFT properties. This illustrates that it is possible to understand some of the properties of the nonlinear response just from the low energy theory, that is, independently from the microscopic details. 
Second, the constraining power of the EFT methods is expected to apply to many other nonlinear elasticity parameter,  beyond the one examined in Ref.~\cite{Alberte:2018doe}. This is especially clear taking into account that the main benefit from the EFT methods is that the full nonlinear structure of the theory is fixed by the symmetries. 

This encourages us to continue the analysis to other cases, in particular to the more sophisticated materials that exhibit {\it scale invariance} (SI). 
As elaborated in \cite{Baggioli:2019elg}, this case deserves a special treatment, because SI can be realized in several ways and this affects the elastic response -- even in the linear response regime.\footnote{We shall not attempt to identify which materials accomplish such a feat. See e.g. \cite{Boyle:2018uiv} for a recent discussion of the possible realization of conformal symmetry in quasi-crystals and \cite{PhysRevResearch.2.022022} for a possible relation between quasi-crystals and our holographic models.}
The main distinction concerns whether SI is a broken or a manifest symmetry of the low energy dynamics. The latter case implies that the theory that describes the solid excitations must be akin to a conformal field theory (CFT). In this case, bottom-up  AdS/CFT methods \cite{Baggioli:2019rrs} provide a useful tool to properly model the material. The opposite case -- with SI as a spontaneously broken -- can be treated instead within the more conventional EFT methods \cite{Leutwyler:1996er,Dubovsky:2011sj,Nicolis:2013lma,Nicolis:2015sra}. 

The existence of these two types of SI solids gives a `bonus' of motivation to the present study, as it is interesting to compare how much the low-energy constraints in the nonlinear response differ depending on whether SI is manifest or spontaneously broken. 
As we will see, there is a significant difference in the relation between the nonlinear parameters introduced above  ($\nu$ and  $\epsilon_{max}$)  for the two types of SI materials.

A good fraction of this work is devoted to provide the tools to compute the nonlinear mechanical response for the manifest SI case, by exploiting  holographic AdS/CFT methods.
The main technical development that we present is the construction of a large family of asymptotically AdS black brane solutions that are subject to finite mechanical deformations\footnote{Let us stress that the elastic response exhibited by our solutions differs from other notions of elasticity black brane horizons discussed previously \cite{Emparan:2009at,Emparan:2016sjk}.} and we obtain their stress-strain curves.
We shall find that a certain class of models allows for black branes that can be deformed elastically by large amounts without breaking. In these cases, their stress-strain diagrams are similar to that of natural rubber (with $\mathcal{O}(1)$ values of the exponents $\nu$, see below), so isn't a great stretch to call these solutions {\it black rubber}.

These solutions can be found semi-analytically in the simplest models, which include the displacement vectors $\pi^i$ as new explicit degrees of freedom -- also called St\"uckelberg fields in the previous literature.  
This paper builds on the recent holographic {\it massive gravity} models \cite{Vegh:2013sk,Blake:2013owa,Andrade:2013gsa,Baggioli:2014roa,Alberte:2015isw,Ammon:2020xyv} which realize in a simple way the spontaneous breaking of translational invariance in `critical' materials (with manifest SI at low energies). 

More recently, several works have improved the framework to accommodate for the spontaneous breaking of translations  \cite{Baggioli:2014roa} and the study of the linear elastic response \cite{Alberte:2015isw}, the vibrational modes of the systems \cite{Alberte:2017cch,Alberte:2017oqx}, their viscoelastic nature \cite{Alberte:2016xja,Baggioli:2018bfa,Andrade:2019zey,Baggioli:2019mck} and their hydrodynamic and physical description \cite{Ammon:2020xyv,Baggioli:2020nay}.

\section{Nonlinear Elastic response}\label{sec:elastic}
In this Section, we review the basic formalism to describe the elastic response under finite (``large'') deformations or applied stresses.  
Linear elasticity theory describes how materials deform in presence of a small (``infinitesimal'') external deformation. The mechanical deformation for a solid in $d+1$ dimensions can be described by a $d$-dimensional vector field, the \textit{displacement vector},
\begin{equation}
\pi_i(x)\,,
\end{equation}
that measures the physical distance from equilibrium position at any given point in the solid.  Out of the $\pi_i$, one builds a rank-2 symmetric tensor, called  the \textit{strain tensor} as
\begin{equation}
    \epsilon_{ij}\,\equiv\,\partial_i \pi_j\,+\,\partial_j\,\pi_i~.
\end{equation}
Volume-preserving deformations satisfy 
\begin{equation}\label{infshear}
\epsilon_{ii}=0
\end{equation}
and are called {\it shear strain}. Similarly, strains that change volume but not shape satisfy 
\begin{equation}\label{infbulk}
\epsilon_{ij}\propto \delta_{ij}
\end{equation}
 and they are called \textit{pure bulk deformations}. 

In most materials, at small enough deformations (small strains), there is a linear relation between the stress needed to apply on the material and the generated strain. Mathematically, this translates into a linear relation between $\epsilon_{ij}$ and the \textit{stress tensor} $\sigma_{ij}$ of the form,
\begin{equation}
\sigma_{ij}\,=\,C_{ijkl}\,\epsilon_{kl}~.
\end{equation}
The  elastic tensor, $C_{ijkl}$, is well known to reduce to just two parameters for homogeneous and isotropic materials: the shear and bulk moduli, which encode the linear response to pure shear and pure bulk deformations respectively.

Non-linear elasticity concerns the relation between stress and strain beyond linear level -- conceptually, the full functional form of $\sigma_{ij}=\sigma_{ij}\left(\epsilon_{kl}\right)$.
In order to extend the relation to the non-linear regime, one must pay attention to how the strain deformations are defined nonlinearly. 

Reviewing the logic, one realizes that in materials that are homogeneous and isotropic at long wavelengths there symmetry allows to choose what we call solid elements so that their equilibrium positions coincide with the `cartesian' coordinates. This suggests to introduce another variable to describe the state of deformation,
\begin{equation}\label{Phi}
\phi^I= \delta^I_i x^i  +\pi^I\,,
\end{equation}
so that equilibrium corresponds to ${\phi}^I_{\text{eq}} = \delta^I_i x^i$. This variable is also more amenable to treat homogeneity and isotropy as an internal symmetry for the scalar fields $\Phi^I$, as done in \cite{Leutwyler:1996er,Dubovsky:2011sj}, which is why the index label on $\pi^I$ has been capitalized. 

A general state of deformation that is constant along the material is then given by
\begin{equation}\label{finitedef}
\phi^I=O^I_j\,x^j\;,
\end{equation}
with an arbitrary constant matrix $O^I_j$, which is a useful way to parameterize the strain tensor for finite deformations. One can easily convince oneself that in the homogeneous and isotropic limit one can restrict $O^I_j$ to be a symmetric matrix with no loss of generality. Isotropy forbids the presence of any background shear strain.

The advantage of using \eqref{Phi} as a variable is now clear: the natural extension that supersedes \eqref{infshear} to the nonlinear regime is
\begin{equation}
{\rm Det}\left( \,O^I_j \,\right)=1\;.
\end{equation}
This condition extends to non-linear level the requirement that the deformation described by the matrix $O^I_j$ does not change the volume of the system.\\

For illustration, in $2+1$ dimensions,  $O^I_j$ is simply a $2\times2$ symmetric matrix, which contains only 3 free parameters. We shall stick to the following parametrization,
\begin{equation}\label{OIJ}
O^I_j\,=\,\alpha\begin{pmatrix} 
\sqrt{1+\varepsilon^2/4} & \varepsilon/2 \\
\varepsilon/2 & \sqrt{1+\varepsilon^2/4} 
\end{pmatrix}~,
\end{equation}
where $\varepsilon$ serves as a nonlinear version\footnote{The volume-preserving nonlinear shear-strain denoted by $\varepsilon$ should not be confused with the strain tensor, which we denote as $\epsilon_{ij}$. } of the shear deformation and $\alpha$ for the pure bulk deformations. We are dropping the `third' parameter, $\tilde\varepsilon$, for deformations of the form $O^I_j={\rm diag}(e^{\tilde\varepsilon},e^{-\tilde\varepsilon})$ because they only differ from $\varepsilon$ deformations in that the shear is introduced in a basis rotated by $45$ degrees. Since we are assuming homogeneous and isotropic materials, it suffices to consider one of the two shear `polarizations'.

The stress-strain curves can now be extracted by computing the stresses in the material that are necessary to support a configuration \eqref{finitedef}. Once the low energy theory for the material is specified, this reduces to just looking at the stresses produced by these configurations. Continuing in the $2+1$ example above, this can be done by computing the stress 
\begin{equation}
    \sigma_{ij} \equiv T_{ij}
\end{equation} 
as a function of the deformation $O^I_j$.

Materials that can be deformed by large amounts while in a reversible fashion are generically called \textit{hyper-elastic}. For these, the stress-strain relation can be obtained from a so-called {\it energy function} scalar function  \cite{enlighten70333}, 
\begin{equation}
    {\cal E}(O^I_{j})\,,
\end{equation}
that characterizes how energetically `expensive' every deformation is. 

From now on, we will consider materials with this property and which realize scale invariance (SI), and we will 
distinguish between two sub-cases depending on how SI is realized: 
\begin{itemize}
    \item[i)] {\it critical} solids, that is, which realize SI as a manifest symmetry at low energies (Section \ref{section:setup});
    
    \item[ii)] solids with spontaneously broken SI, with  a gapped spectrum and thus have phonons as the lowest-energy excitations (Section \ref{section4}).
\end{itemize}
Case ii) can be dealt with using the solid EFTs, so part of the discussion was already presented in \cite{Alberte:2018doe}. Here we will extend the analysis with the aim at the SI case and its  comparison  to the manifest SI case.

As a warm-up, let us remind now how the computation of the stress-strain curve proceeds for a general solid EFT (not necessarily assuming broken SI).
Restricting to 2+1 dimensions for simplicity, it can be seen that the most general effective Lagrangian at leading order in derivatives  can be written as \cite{Nicolis:2013lma,Alberte:2018doe}
\begin{equation}\label{action}
S= -\int d^3x\,\sqrt{-g}\, V(X,Z)\,,
\end{equation}
with $X$ and $Z$ defined in terms of the scalar fields matrix\footnote{We retain the curved spacetime metric $g_{\mu\nu}$ only to make it clear how the energy-momentum tensor  arises from this action. In practice we shall always work on the Minkowski background $\eta_{\mu\nu} = \textrm{diag}\,(-1,+1,+1)$.} 
$\mathcal I^{IJ}= g^{\mu\nu}\partial_\mu\phi^I\partial_\nu\phi^J$ as 
$
X =\frac{1}{2}\,\mathrm{tr }\, \big(\mathcal I^{IJ} \big) \,,
\, Z = \det \big(\mathcal I^{IJ} \big)$.

It is immediately clear that once one restricts to the (strained) static and homogeneous  configurations given by  \eqref{finitedef} and \eqref{OIJ}, the action \eqref{action} plays exactly the same role as the `energy function'. In other words, we can identify
\begin{equation}
V(X,Z)\big|_{\rm constant~strain} = {\cal E}^{{}^{EFT}}(O^I_{j})~.
\end{equation}
We emphasize that the nontrivial content in the Solid EFT construction is that once one knows the `energy function' then the whole effective Lagrangian is also known, which can then be used to extract more information such as the elasticity bounds \cite{Alberte:2018doe}.
Instead, for the solids with manifest SI of Section \ref{section4} the energy function still exists but it does not correspond directly to the effective Lagrangian -- in fact in these cases one expects that a local Lagrangian doesn't exist.

The stress required to support the configurations \eqref{finitedef} can be read off from the stress tensor associated to these configurations \eqref{finitedef}, which can be easily computed in the EFT. 
For any time independent scalar field configurations, the stress-energy tensor components are \cite{Alberte:2018doe}
\begin{align}\label{rho}
&T^{tt}\,\equiv\,\rho\,=\,V\,,\\\label{pressure}
& T^{xx}\,\equiv\,\,p\,=\,-\,V\,+\,X\,V_X\,+\,2\,Z\,V_Z\,,\\\label{txy}
&T_{xy}\,=\,\partial_x \phi^I \partial_y \phi^I\,V_X\,,
\end{align}
where $V_X\equiv\partial V /\partial X,$ etc. 
The deformed field configuration \eqref{finitedef} introduces both shear and bulk deformation. Setting $\alpha=1$, it describes a \textit{pure shear} strain (\textit{i.e.} volume-preserving) in the $(x,y)$ directions induced by $\varepsilon\neq 0$. 

The full nonlinear stress-strain curve is then found be expressing the stress $T_{xy}$ as a function of the strain $\varepsilon$,
\begin{equation}\label{sigma}
\sigma(\varepsilon)\equiv T_{xy}=\,\varepsilon\,\sqrt{1+\frac{\varepsilon^2}{4}} \;
V_X\left(1+\frac{\varepsilon^2}{2},1\right)\,.
\end{equation}

These results apply to any solid whose low energy dynamics can be treated with EFT methods. This includes the solids with spontaneously broken scale invariance (SI), which we discuss in more detail in Section \ref{section4}.
However, these steps are not justified for solids which exhibit manifest scale invariance \cite{Baggioli:2019elg}. 

In principle the procedure is identical for solids with manifest SI, we just want to obtain how much stress is required to support a configuration with given strain $\varepsilon$. %
However, the main obstacle is that, as in CFTs, scale-invariant solids are expected to lack a local Lagrangian description, therefore the  steps after \eqref{action} do not immediately apply (nor the identification of the 'energy function' with an effective Lagrangian). While this may seem unimportant regarding the response to static and homogeneous strain, 
it is crucial in order to possibly  obtain nontrivial constraints in the nonlinear response (such as the correlations among various nonlinear  parameters mentioned in the introduction) because this requires a knowledge of the full theory.

In the next Section, we show how to extract stress-strain curves in (holographic models of) solids with manifest SI, we shall work out the equivalent of Eq.\eqref{sigma} for them, and find the constraints and relations among different nonlinear elasticity observables.

\section{Solids with manifest scale invariance}
\label{section:setup}
As mentioned in the introduction, our main focus are  materials in a critical regime -- which exhibit manifest scale invariance at low energies. We shall model them  using the standard holographic dictionary. As usual, it simplifies the analysis to model the scale invariant field theory as a deformation of a CFT. In this case, the AdS/CFT dictionary tells us that the material, which we assume is 2+1 dimensional, is going to be described by asymptotically AdS${}_4$ planar black brane solutions. By assumption, the CFT contains operators that can be identified with the displacement vectors. Their dual incarnation in the AdS${}_4$, are an identical a set of fields,  $\phi^I$, which propagate into the holographic dimension too. See \cite{Alberte:2015isw,Baggioli:2016rdj} for more details.

The way to extract the stress-strain curves in these models is simply to find the black brane solutions with nontrivial strain tensor `emanating' from the horizon. The strain tensor, then, can be thought of as an asymptotic charge of these black branes. Keeping track of the stress tensor for each strain tensor `charge', one can compute the strain-stress curve.

Given that for every (constant) value of the strain tensor there is a static black brane solution, the process of varying the strain (which is implied in the stress-strain curves) can be assumed to be a reversible process, if done slowly enough. For this reason, we will treat the (static) nonlinear elastic response of these black branes as elastic (reversible). This is, of course, until some instability is reached - and this is the basic guiding principle we shall use to establish elasticity bounds.

\subsection{Nonlinear response for holographic models}

We consider the holographic {\it massive gravity} models introduced in \cite{Baggioli:2014roa,Alberte:2015isw} (see also \cite{Baggioli:2015zoa,Baggioli:2015gsa,Baggioli:2015dwa,Alberte:2016xja,Alberte:2017cch,Alberte:2017oqx}), which are obtained  by introducing displacement fields $\Phi^I$ in the AdS bulk, with a generic action 
\begin{equation}
\mathcal{S}\,=\,\int d^{4}x\,\sqrt{-g}\,\left[\,R-2\,\Lambda-\,2\,m^2\,W(X,Z)\,\right]\label{model}\,,
\end{equation}
with $\mathcal{X}^{IJ}\equiv\partial_\mu \Phi^I \partial^\mu \Phi^J$ and $X\equiv \frac{1}{2}\mbox{Tr}(\mathcal{X}^{IJ})$ and $Z\equiv \det(\mathcal{X}^{IJ})$. For simplicity, we focus on $d=3$ but we the construction can be easily extended to higher dimensions.  

For specific choices of the potential $W(X,Z)$, the model \eqref{model} represents the gravity dual of a CFT at finite temperature and 
%zero charge density 
where translational invariance is broken spontaneously.  Using the standard AdS/CFT dictionary, this defines for us a CFT that will have non-zero elastic moduli and so it can be interpreted as a model for a solid in a quantum critical regime. We remind the reader that throughout all the manuscript we will only consider standard quantization for the scalar bulk fields $\Phi^I$. 
More precisely, under these assumptions, a well-defined elastic response can be defined for potentials which decay at the boundary as $W \sim u^3$ or faster \cite{Alberte:2017cch} \footnote{One could try to avoid this constraint by using alternative quantization; nevertheless, the corresponding models would be dynamically unstable due to a negative shear modulus \cite{Ammon:2020xyv,Baggioli:2020ljz}.}. Moreover, for potentials whose fall-off at the boundary is $W \sim u^5$ or faster this elastic response is associated to the presence of massless propagating phonons \cite{Alberte:2017oqx}. The most important point for the moment is that the gravity theory also contains a field $\Phi^J$, which is directly linked to the material deformation.

It has been shown before \cite{Bardoux:2012aw} that there exist simple homogeneous asymptotically AdS planar black brane solutions with 
\begin{equation}\label{scalarHair}
\Phi^J(u,x)=\delta^J_j \,x^j~,
\end{equation} 
which from the gravitational perspective acts as a ``\textit{solid hair}" or more technically as magnetically charged 0-forms \cite{Caldarelli:2016nni}.
Their CFT interpretation fits that of a critical 2+1 planar and homogeneous solid material, with broken translations.
How to perturb this solution and read-off the (linear) elastic moduli has already been discussed at length previously \cite{Alberte:2017cch,Alberte:2016xja,Alberte:2015isw}. 

Our next goal is to find the holographic stress tensor carried by strained configurations (strained solids)
\begin{equation}\label{stainHair}
\Phi^I(u,x)=O^I_j \,x^j~,
\end{equation} 
with $O^I_j$ given in \eqref{OIJ} (with finite $\varepsilon$ and $\alpha$). Since the CFT stress tensor is dual to the metric and we are after the full nonlinear response, we must find the {\rm exact} holographic stress tensor produced by the deformation `source' $O^I_j$. 

In practice, this implies that we must find (asymptotically AdS black brane) exact solutions to the Einstein $+$ scalars theory with a nonzero tensor mode  -- the strain tensor. That is, the spatial part of the metric $g_{ij}$ (with $i,\,j$ running over $x,\,y$) must differ from $\propto \delta_{ij}$ so  that it contains a shear (and bulk) deformation. 

Fortunately, for deformations that that are constant in time and space it is possible to reduce significantly the equations \footnote{For a more complicated case of oscillatory shear deformations in the same class of models see \cite{Baggioli:2019mck}.}.
Indeed, one can see that the full system of nonlinear Einstein equations can be solved in this case going to  the following ansatz 
\begin{equation}
ds^2\,=\,\frac{1}{u^2}\left(-f(u)\,e^{-\chi(u)}\,dt^2+\frac{du^2}{f(u)}+\gamma_{ij}(u)\,dx^i dx^j\right)\,,\label{geometry}
\end{equation}
where $\gamma_{ij}$ is a $d-1$ dimensional spatial metric with unitary determinant. In $d=3$, one can parametrize  $\gamma_{ij}$ in terms of the usual $+$ and $\times$ polarizations as
\begin{equation}
\widehat{\gamma} \, = \, {\rm exp}\left[ h_{+}(u)\,\widehat{\sigma}_{+} \,+\, h_\times(u)\,\widehat{\sigma}_\times \right]\,, 
\label{nonlindef1}
\end{equation}
where $h_{+,\,\times}$ are functions of $u$ only and  $\widehat\sigma_{+,\,\times}$ stand for the Pauli matrices that are usually called $\sigma_{3,\,1}$ respectively. 

The two polarizations $h_{+,\times}$ couple to each other at nonlinear level. In order to disentangle them, it is convenient to switch variables to 
\begin{equation}
h_\times = h \cos\theta
\qquad 
h_+ = h \sin\theta\,,
\label{h&theta}
\end{equation}
where again $h$ and $\theta$ are functions of $u$ only. 
In these variables, $h$ is the magnitude of the spin-2 mode and $\theta$ the direction in the space of polarizations. 
For physical solutions, $h(u)$ must have a vanishing leading mode - and its subleading mode encodes the  stress tensor.
As we will see shortly, it follows from the equations of motion that in these solutions the function $\theta(u)$ must be a constant. 
In these variables, then,  $\theta$ will simply encode the polarization direction of the stress tensor and $h$ the magnitude of the response.

We can do a similar representation for the strain matrix $O^I_j$:
\begin{equation}\label{scal}
%\phi^I \, = \, \xi^I_j \,x_k\, \delta_{jk}\,, \hspace{0.7cm} \mbox{where} \hspace{0.7cm} 
\widehat{O}\, =\alpha \, \; {\rm exp}\;{\left[\,\frac{\Omega}{2} \,\, 
\big(\cos\theta_0 \; \widehat\sigma_\times
+\sin\theta_0 \; \widehat\sigma_+\big) \right]} ~.
\end{equation}
The constant $\alpha$ parametrizes the bulk strain deformation whereas the constants $\Omega$ and $\theta_0$ encode the strain magnitude and polarization, and they are related to the nonlinear shear strain parameters $\varepsilon$ and $\tilde\varepsilon$ introduced in Section~\ref{sec:elastic}. For instance, for $\theta_0=0$, one has
\begin{equation}
    \varepsilon=2\sinh\left(\Omega/2\right)~.
\end{equation}
The magnitude of the shear strain encoded in $\Omega$ acts as a source term for the metric in the bulk.

Before showing the equations of motion, note that in homogeneous and isotropic material the elastic response is such that the strain and stress tensors are aligned in the same polarization direction. In our notation, this translates to having $\theta=\theta_0$. We shall see shortly that this is indeed the case in for the physical solutions, but for the moment we keep $\theta(u)$ generic in order to see how it is determined by the equations of motion.\\

In $d=3$, the independent equations for the background \eqref{geometry} are:
\begin{align}%\label{eqs}
& 2\,\chi'\, -\,u\,\left(\sinh^2(h)\,\theta\,'^{\,2}\,+\,h'^2 \right)\,=\,0,\label{chieq}\\[3mm]
&u\,f'-\Lambda\,-\,m^2 \,W(\bar{X},\bar{Z})-(6+u\,\chi')\,f/2 = \, 0\,, \\[3mm]
& f\,\left(\,2\,u^2\,h''\,-\,u^2\,\sinh (2\,h)\,\theta\,'^{\,2}
- u\, h'\,(4+u\,\chi')\,\right)\,+\,2\,u^2\,f'\,h'\nonumber\\[3mm]
&\,-\,4\,m^2\,W_h (\bar{X},\bar{Z})\,=\,0\,,\\[3mm] 
&f \,\left(\,2\,u^2\,\theta\,''+4\,u^2\,\coth(h)\,\theta\,'\,h'-u\,\theta\,'\,(4\,+\,u\,\chi')\,\right)\,+\,2\,u^2\,f'\,\theta\,'\,\nonumber\\[3mm] 
&\,-\,4\,m^2\,W_{\theta} (\bar{X},\bar{Z}) \, {\rm cosech}^{2}(h)\, = \, 0 \,, \label{thetaeq}
\end{align}
where the cosmological constant is fixed to $\Lambda=-3$ and we indicate with the subscript $(h,\theta)$, the derivative with respect to $h$ and $\theta$. 
The potential $W$ is evaluated on the background values, which gives
\begin{align}\label{back}
\bar{Z}\,\equiv  \, \alpha ^4 u^4 \,, \qquad
\bar{X}\,\equiv & \, \alpha ^2 u^2 (\cosh{\Omega}\;\cosh{h}\,-\,\cos(\theta_0 -\theta)\;\sinh{\Omega}\;\sinh{h}) .\nonumber
\end{align}

Within this ansatz the form of $\chi(u)$ and $f(u)$ are completely dictated by  $h(u)$ and $\theta(u)$. We assume the presence of an event horizon at $u=u_H$ defined by $f(u_H)=0$ and $h(u)$ reaching a constant value at the horizon, $h(u_H)=h_H$. The associated entropy density is $s=2 \pi /u_H^2$ and the corresponding temperature reads $T=-\frac{f'(u_H)}{4\,\pi}\,e^{-\chi(u_H)/2}$. In the asymptotic UV region we impose $f(0)=1$ and $\chi(0)=0$.

Assuming that the mass term $m^2 \,W_h$ vanishes sufficiently quickly towards the asymptotic UV region, one finds that the two independents modes of the spin-2 metric deformation are
\begin{equation}
h(u)\,=\,\mathcal{C}_0\,\left(1\,+\,\dots\right)\,+\,\mathcal{C}_3\,u^3\,+\,\dots \label{exp}
\end{equation}
where dots represent higher powers of $u$. As usual, the subleading term $\mathcal{C}_3$ is identified via the AdS/CFT dictionary with the VEV of the stress tensor $ \langle T_{xy} \rangle$, and $\mathcal{C}_0$ with an external spacetime metric source for $T_{xy}$ operator (see below). 
Throughout all the manuscript, we will consider that the external spacetime deformation source is absent, so that
\begin{equation}
\mathcal{C}_0\,=\,0\,.
\end{equation}
With this condition, the only possible source for the stress tensor arises from the mechanical strain deformation that is encoded in the scalars $\Phi^I$ -- ultimately in the parameters $\alpha$,  $\Omega$ and $\theta_0$ that parametrize the strain deformation.

We can now look at the equation for $\theta(u)$, \eqref{thetaeq}. Due to the cross-coupling to $h$ in the second term of   \eqref{thetaeq}, the two modes of $\theta(u)$ near the AdS boundary turn out to depend on the boundary condition assumed for $h$, {\it i.e.} on the choice of $\mathcal{C}_0$. For  $\mathcal{C}_0\neq0$,  $\theta$ would have a constant mode and a  $u^3$ mode. However, for $\mathcal{C}_0=0$, the asymptotic $\theta$ modes are the constant mode and a $u^{-3}$ mode. Regularity at the AdS boundary (rather, consistency with AdS asymptotics) then requires to set the $u^{-3}$ coefficient to vanish. Given that one is limited to only one free parameter, regularity at the horizon then is then expected to select the $\theta(u)=\theta_0$ as the only viable solution at least for generic choices of the potential $W$. Incidentally, this closes the proof that the elastic response is isotropic, since a strain in a given polarization only sources stress tensor in the same polarization (this would not be true if $\theta(u)$ had a nontrivial profile). This was completely expected, but it is easy to show expliticly in the variables \eqref{h&theta}.

Therefore, from now on we will set $\theta=\theta_0 = 0$ for the rest of the manuscript. This simplifies the set equations of motion substantially, 
\begin{align}\label{simpler}
& 2 \left(\,u\,f'- \Lambda\,-\,m^2 \,W(\bar{X},\bar{Z})\, \right) 
- f\, \left(6+u^2\, {h'}^2/2\right)\,=\,0\,,\nonumber\\[3mm]
&f\,\left(2\,u^2\,h''-u\,h'\,(4\,+\,u^2\,{h'}^2/2)\right)+2\,u^2\,h'\, f'\,-4\,m^2 \,W_h(\bar{X},\bar{Z})=0,
%\nonumber \\
% &  2\,\chi '\,=\,u\, {h'}^2,
\end{align}
where now we can write
\begin{equation}
    \bar{X} \equiv u^2\alpha^2 \cosh (h-\Omega), 
\end{equation}
which makes clear that $\Omega$ acts as a `source' term for $h$ in the bulk. 
These two equations \eqref{simpler} determine uniquely the profiles for $f$ and $h$ and then $\chi(u)$ is obtained by integrating \eqref{chieq}, which reduces to $2\,\chi '\,=\,u\, {h'}^2$.

\subsection{General results}

From the point of view of the gravity theory in the bulk, the solutions to \eqref{simpler} can be viewed as black branes with a form of hair  encoded in the scalar configurations with nontrivial strain \eqref{OIJ}. To fix ideas, we can think that there are 2 parameters (or charges) that label the solutions: the magnitude of the scalar gradient at zero shear strain, $\alpha$, which we assume is nonzero (and which can be traded by $m$ for monomial potentials);
and the magnitude of the shear strain $\varepsilon$. (We ignore now the angle $\theta_0$ since it only sets a direction.)
The novelty of the solutions presented here with respect to the ones previously discussed {\it e.g.} in \cite{Baggioli:2014roa} is that we will keep track of how the finite
shear strain $\varepsilon\neq0$ deforms the solutions. 

Picturing the shear strain as a standard charge that the black branes can be endowed with is also useful to understand their behaviour and properties. 
It is clear from \eqref{geometry} that we are constructing solutions with a non-zero and {\it static} tensor mode of the metric, $h(u)$. 
This is possible for two reasons: first, because the strain tensor encoded in the scalars acts as a source for the tensor mode $h(u)$; second, because the tensor mode $h$ is a {\it massive graviton}. Indeed, the presence of a mass term in the equation of motion grants the possibility to have static response to a static homogeneous source.

At the level of understanding the stress-strain curves that will follow, it is clear that increasing the shear strain one must reach extremal ($T=0$) solutions. Also, by changing $u_H$ together with $\varepsilon$, it is possible to construct one-parameter family of solutions, say, at constant temperature. Labeling these solutions by the amount of strain $\varepsilon$, and computing the shear stress for each solutions then we can obtain the strain-stress curve.  This is the strategy that we follow in this work. Let us now summarize two general results that follow from this prescription.

First of all, it is possible to obtain an approximate expression for the stress-strain curve implied by  \eqref{simpler} for shear deformations (that is with $\alpha=1$). The main observation is that the $m^2$ term in \eqref{simpler} acts as a source term for $h$ for $\Omega\neq0$ and that at either $m=0$ or $\Omega=0$,  $h(u)=0$ is a solution -- which means that the stress vanishes in this limit.  
The profile $h(u)$ is guaranteed to be small then for small $m$ (at least for a class of potentials), and this allows for a perturbative scheme even for large deformations, $\Omega\gtrsim1$ (equivalently,  $\varepsilon\gtrsim1$). 
Following \cite{Alberte:2016xja}, we can treat $m$ as a small parameter and find the solution order by order in $m^2$.
At first order in $m^2$ this gives
\begin{equation}
\sigma(\varepsilon)=\frac{1}{2}\,m^2\,\varepsilon \, \sqrt{4+\varepsilon^2} \int_0^{u_H}\,
\frac{W_X\left(\frac{1}{2}\,(2+\varepsilon^2)\,\zeta^2,\zeta^4 \right)}{\zeta^2}\,d\zeta \;+\; {\cal O}(m^4)\,.
\label{integral}
\end{equation}
This formula is of course only applicable if the integral is finite, which translates into some constraints on the functional form of the potential $W(X,Y)$. We will elaborate more on the constraints on $W(X,Y)$ in the coming subsections. 
At this stage, let us make two remarks. First \eqref{integral} is perfectly consistent with the 
perturbative expression for the shear modulus  found previously in \cite{Alberte:2016xja}.  Second, the formula \eqref{integral} resembles structurally the one arising from EFT methods \eqref{sigma} but it also differs substantially in various ways: the bulk potential enters inside an integral which hints at a sense of non-locality; additionally, \eqref{integral} encodes a non-trivial temperature effect, via the dependence on the location of the horizon, $u_H$ (which was obviously absent in the EFT formalism). 
 
Since the expression \eqref{integral} is obtained perturbatively in $m^2$, the factors next to $m^2$ should be not too large in order to be valid and at large enough  $\varepsilon$ one expects that this approximation should fail. However, as we will see below the \eqref{integral} still gives a good approximation even for moderately large $\varepsilon$.

The form of the stress-strain curve at very large strain can still be obtained from a different consideration.  
The behaviour at asymptotically large strain can be understood  from the structure of  equations of motion \eqref{simpler}. The key point is that large strain implies $h(u)\gg1$ and in this very anisotropic regime the equations \eqref{simpler} have an  attractor solution (towards the UV), different from $AdS_4$. In a subset of models (defined by the choice of the potential $W(X,Z)$), this UV attractor solution is actually a fixed point that realizes scale invariance with anisotropic scaling in the spatial $x\,,y$ directions as well as in time.
The presence  of this additional anisotropic Lifshitz UV fixed point  translates into the appearance of a second  power-law scaling behaviour at asymptotically large  strains. For the sake of clarity, we postpone the discussion of this point to Sec.~\ref{section:nonlinear} in the context of a specific benchmark model.

Finally, the models \eqref{model} admit
extremal solutions of the form \eqref{geometry}, whose near-horizon geometry are then  $AdS_2\times \mathbb{R}^2$. This represents yet another additional emergent scale invariance, this one with with Lifshitz dynamical exponent $z\to\infty$, and isotropic character. This scaling is expected to be manifested in the lightest excitations, governed by the near horizon geometry.

\subsection{A benchmark model}\label{section:nonlinear}

In order to make further progress, the form of the potential $W$ must be specified. In this work we shall not  try to find what is the form that matches the mechanical response of some known materials. Rather, we shall take an approach similar to \cite{Baggioli:2019elg}, where we concentrate on potentials $W$ that give rise to power-law stress-strain relation $\sigma\sim\varepsilon^\nu$.

For this purpose we consider a benchmark potential of the form:
\begin{equation}
W(X,Z)\,=\,X^\mathfrak{a}\,Z^{\frac{\mathfrak{b}-\mathfrak{a}}{2}}\,.\label{bench}
\end{equation}
In order to ensure the consistency of this choice \eqref{bench}, and of the model \eqref{model} in general, we ask the following requirements: absence of ghosts, absence of gradient instabilities, and positivity of the linear elastic moduli in the backgrounds with vanishing shear stress, $\varepsilon=0$. These conditions constrain the range of the parameters $\mathfrak{a},\mathfrak{b}$ in the benchmark model \eqref{bench} as follows  \cite{Baggioli:2019elg},
\begin{equation}\label{cond}
\mathfrak{a}\geq 0,\,\mathfrak{b}\,>\,\frac{3}{2}~.
\end{equation}
Moreover, restricting the discussion to the case of standard quantization, we need to impose also  $\mathfrak{b}\,>\,\frac{5}{2}$ to ensure the presence of massless phonons.

Notice that the constraints considered are mostly bulk requirements and they represent just necessary but not sufficient conditions for the full consistency of our boundary field theory. In order to have a final verdict a detailed QNMs computation would be needed. For simple enough theories (of the form $W(X)=X^n$ or $W(Z)=Z^m$), that has been done in \cite{Alberte:2017oqx,Ammon:2019apj,Baggioli:2019abx,Ammon:2020xyv}.

At low temperatures, we will find two distinct regimes  with a scaling relation $\sigma\sim \epsilon^\nu$ (both for shear and bulk strain deformations), as seen {\it e.g.} in Fig.~\ref{shearfig1}. 
For the sake of clarity, then, we will introduce the following notation for the corresponding exponents: 
$$
\nu_{1}^{S}\,,\quad
\nu_{2}^{S}\,,\quad
\nu_{1}^{B}\,,\quad 
{\rm and} \quad
\nu_{2}^{B}\,,\quad
$$ 
where the $1,\,2$ subscript denotes the regime encountered at lower and higher deformation regimes respectively; and $S\,, B$ stand for the pure shear and pure bulk deformation sectors.

\subsubsection*{Shear deformations}
The nonlinear shear response is encoded in the shear stress strain curve $\sigma(\varepsilon)$, where the stress is given by (see Appendix \ref{appendix:stress}):
\begin{equation}
T^{xy}\,=\,\sigma\,=\,\frac{3}{2}\,\mathcal{C}_3\,,
\end{equation}
and $\mathcal{C}_3$ is the subleading term in the UV expansion \eqref{exp}. The strain on the contrary is produced by the difference between the background configuration \eqref{scal} and the equilibrium one $\Phi^J=x^J$ and it reads:
\begin{equation}
\varepsilon\,=\,2\,\varepsilon_{xy}\,=\,2\,\sinh \left(\Omega/2\right)\,.
\end{equation}
In simple words the shear stress-strain relation is derived by the identification:
\begin{equation}
\sigma(\varepsilon)\,\quad \longleftrightarrow \quad \frac{3}{2}\,\mathcal{C}_3\left(2\,\sinh \left(\Omega/2\right)\,\right)\,,
\end{equation}
where $\mathcal{C}_3$ is extracted numerically varying $\Omega$.

The solutions can be easily obtained by shooting method, integrating \eqref{simpler} from the horizon towards the UV boundary. The boundary conditions that ensures regularity at the horizon are $\left[2\,u^2\,h'\, f'\,-4\,m^2 \,W_h(\bar{X},\bar{Z})\right]_{u_{H}}=0$ and $h(u_H)=h_H$, a finite constant that is used as the shooting parameter.

\begin{figure}[t]
\begin{center}
\includegraphics[height=7cm]{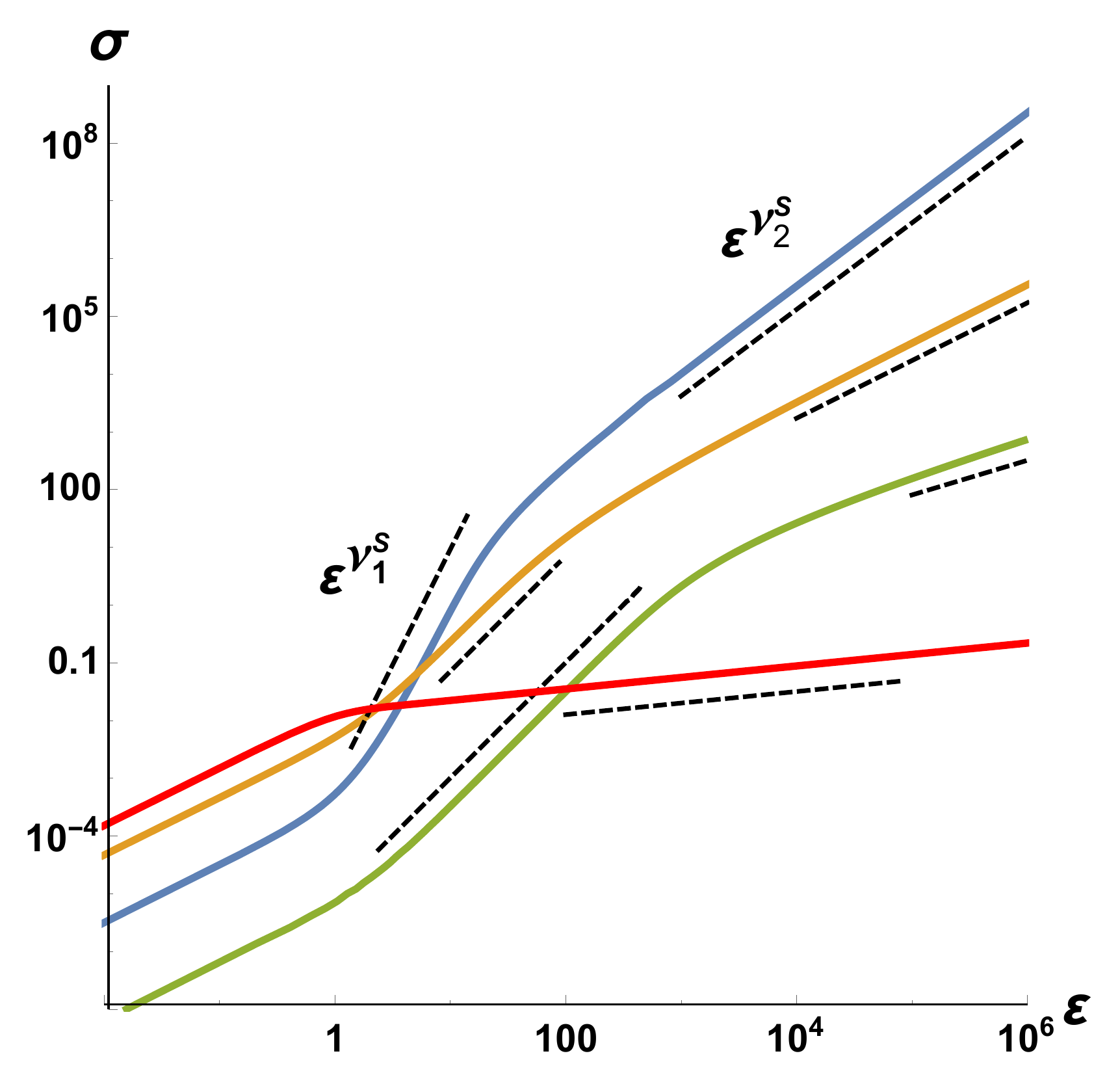}
\quad
\includegraphics[height=7cm]{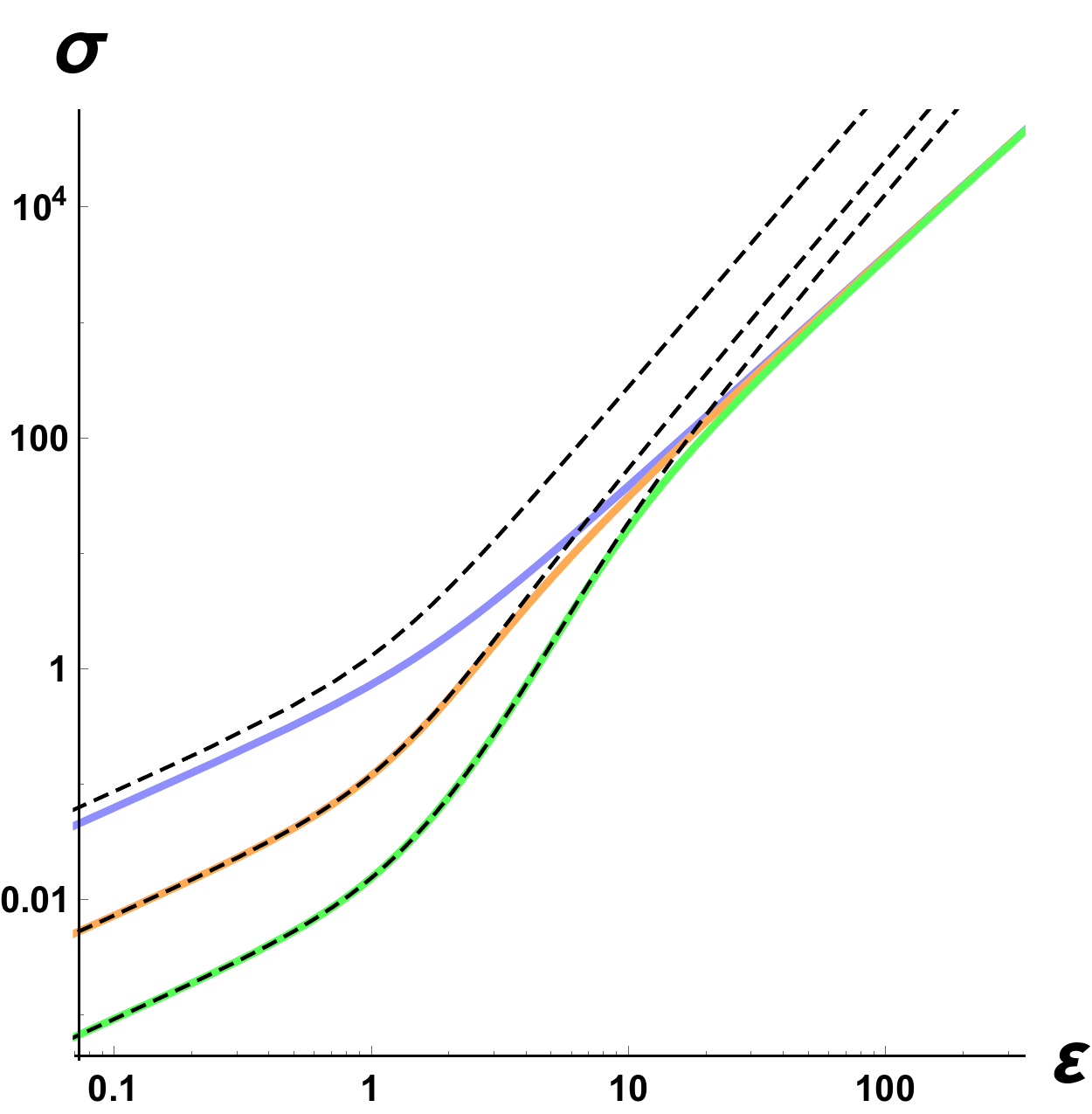}
\caption{\textbf{Left: }Shear stress strain curve for various potentials and relative (dashed) large strain scaling at $m/T=1$. The potentials are chosen to have both stiffening and softening, \textit{i.e.} $\nu_1^S=2\mathfrak{a}=4,2,2,0.2$ and $\nu_2^S = 3\, \mathfrak{a}/\mathfrak{b}=3/2,1,3/5,1/7$. Here the constraints on the strain are not taken into account. \textbf{Right: }Shear stress strain curves of a potential $W(X,Z)=X^2\sqrt{Z}$ for different temperatures ($T/m = 0.1,0.5,1$ for blue, orange and green) and comparison with the analytic formula \eqref{integral} (dashed lines). As expected for $T/m \gg 1$ the formula gives a very good approximation.}
 \label{shearfig1}
\end{center}
\end{figure}

The results for the non linear shear response are shown in Fig.\ref{shearfig1} for an illustrative choice of potentials. Clearly, at small strains $\varepsilon \ll 1$, the response is linear and the slope is given by the shear modulus
studied previously \cite{Alberte:2016xja}. Moving away from the linear approximation, we can notice that the stress-strain curve exhibits two different scaling regimes. At intermediate strain, a power law behaviour $\sigma \sim \varepsilon^{\nu_1^S}$ appears for $\varepsilon\gg1$ (but not too large), with 
\begin{equation}
\nu_1^S = 2\mathfrak{a}~.
\label{nuscal1}
\end{equation}
Additionally, for much larger strains ( $\varepsilon \gtrsim 10^2 - 10^3$ in the examples shown in Fig.~\ref{shearfig1}), the curve again displays a scaling $\sigma\sim\varepsilon^{\nu_2^S}$ with a different exponent
\begin{equation}
\nu_2^S\,= \,3\,\frac{\mathfrak{a}}{\mathfrak{b}} ~.
\label{nuscal}
\end{equation}
The manifestation of two scaling regimes actually happens only at high enough temperature; at low temperature the curve interpolates from the linear regime directly to the $\nu_2^S$ scaling as shown in Fig.\ref{shearfig1}. As shown later, both scalings can be obtained analytically (see formula \eqref{integral}).\\

Let us now supplement the numerical results shown in Fig.~\ref{shearfig1} with some analytical understanding.
Performing the integral in  \eqref{integral} for our benchmark potential \eqref{bench}, we find the approximate expression 
\begin{equation}
\sigma(\varepsilon)\simeq\frac{\mathfrak{a}}{2^\mathfrak{a}\, (2\mathfrak{b}-3)}\,m^2\,\alpha^{2\mathfrak{b}}\,u_H^{2\mathfrak{b}-3}\,\varepsilon\, \sqrt{4+\varepsilon^2}\,(2+\varepsilon^2)^{\mathfrak{a}-1}~.
\label{sigmamT}
\end{equation}
Convergence of the integral requires $\mathfrak{b}>3/2$, the same condition of the positivity of the linear bulk modulus. The linear limit $\varepsilon \ll 1$ of formula \eqref{integral} agrees perfectly with what was found previously in \cite{Alberte:2016xja}.

This is an approximate expression at leading order in $m^2$, therefore it is valid so long as all factors are of not too large, which allows for finite but moderate $\varepsilon$. Still, the large $\varepsilon$ limit of \eqref{sigmamT} already catches the first scaling behaviour with exponent $\nu_1^S$ given in \eqref{nuscal1}. At larger strain, though,  Eq.~\eqref{sigmamT} is not expected to hold and  more effort is needed to understand what should happen.

At asymptotically large strain, $\varepsilon \gg 1$, it is still possible to obtain the non-linear scaling at large strain analytically by analysing the equations of motion \eqref{simpler}.
Large strain implies $\Omega\gg1$ and so the tensor mode profile $h(u)$ is expected to perform a large excursion or, equivalently, to have sizeable gradient $h'(u)$. 
It is convenient to introduce 
\begin{equation}
\tilde h(u) \equiv \Omega - h(u)
\end{equation}
because the potential depends on $h(u)$ only through this combination. At large $\Omega$, near the boundary this quantity is large, and moreover one can expect that $\tilde h'(u)$ is also large (say, compared to $f'(u)$) somewhere in the bulk.

\begin{figure}[t]
\center
\includegraphics[width=9cm]{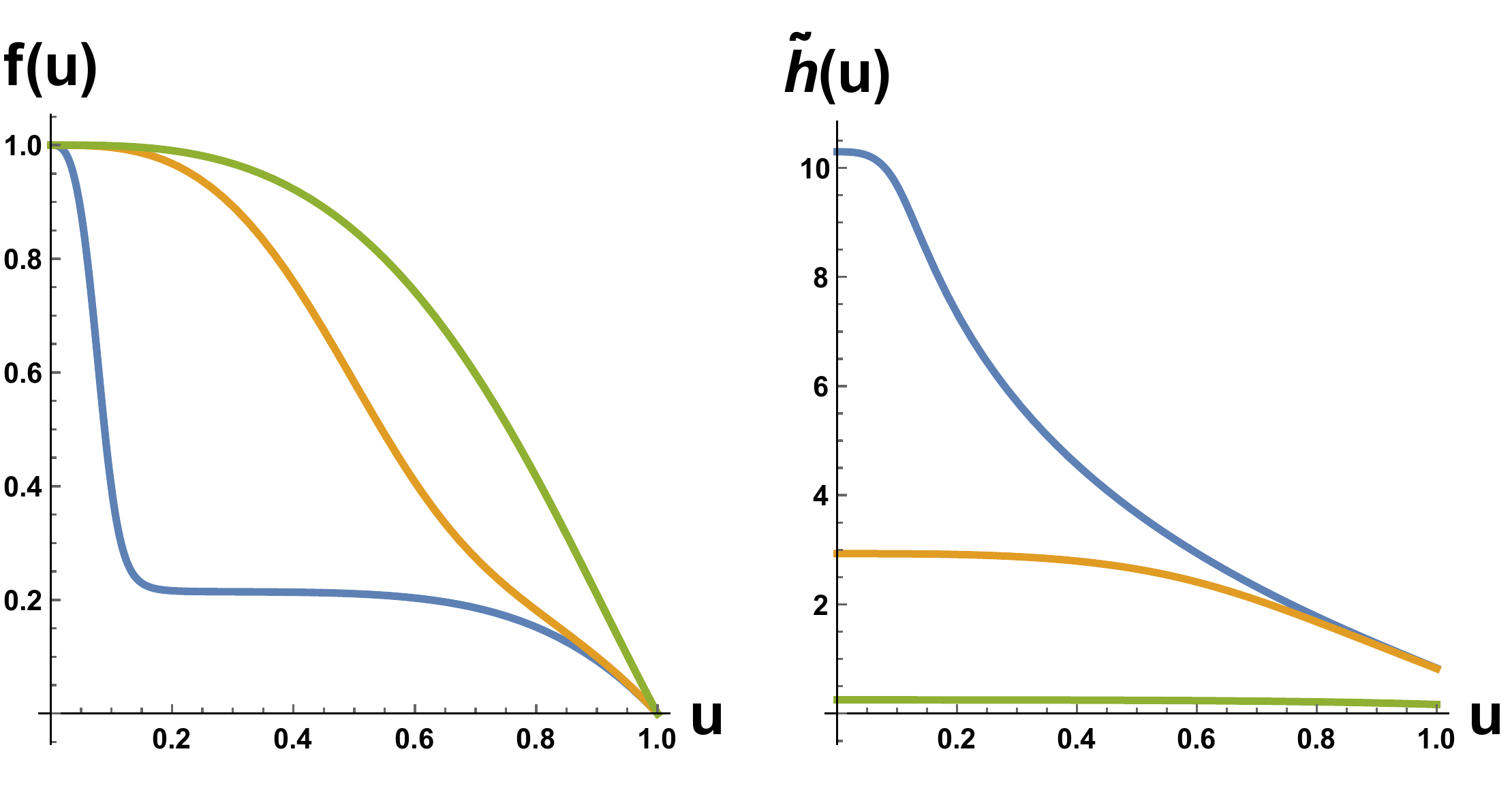}
\caption{The metric functions $f(u)$ and $\tilde{h}(u)$ for a potential $W(X,Z)= X^2 Z$, $u_H=1$, $\Omega = 0.24,\, 2.93,\, 10.29$  and temperatures $T/m \sim 0.15,\, 0.02,\, 10^{-5}$ (Green, orange, blue).}
\label{fandc}
\end{figure}

The large $\tilde h \gg 1$ approximation allows to  substitute $\cosh(\tilde h)$ and $\sinh(\tilde h)$ by $e^{\tilde h}/2$. If the potential $W(X,Z)$ is power-law-like as in our benchmark model, then it is easy to see that in this regime that the solution to the equations of motion  \eqref{simpler} reaches a constant for  $f(u)\simeq f_0 <1$. 
In fact, our benchmark potential \eqref{bench} admits a simple approximate solution of the form near the UV ($u\to0$),
\begin{equation}\label{anis}
f(u)\,\simeq\, f_0 = \frac{3 \, \mathfrak{a}^4}{\left(\mathfrak{a}^2+\mathfrak{b}\right) \left(3 \,\mathfrak{a}^2+\mathfrak{b}^2\right)} \;, \qquad  \; 
\tilde{h}(u)\, \simeq\,  - 2\,\frac{\mathfrak{b}}{\mathfrak{a}}\, \log \left(\frac{u}{u_0}\right)
\;,
\end{equation}
with $m^2 \; u_0^{2b} = 3\,b\,2^a\,/(a^2+b)$. The  logarithmic shape of $\tilde h(u)$ is clearly seen  in Fig. \ref{fandc}. 

Before discussing further the properties of this approximate solution, let us first see how it determines the elastic response at asymptotically large strain. In this limit, the function $\tilde{h}(u)$ at the UV is large but finite, $\tilde{h}(0)=\Omega$. Therefore, at some scale $u_*$, there must be a transition between the constant and the logarithmic profile \eqref{anis}. 
Assuming that \eqref{anis} is correct up to the horizon, 
we have that approximately 
$\tilde{h}(u)\simeq \Omega - 2\,\frac{\mathfrak{b}}{\mathfrak{a}}\, \log \left(\frac{u}{u_*}\right)$ from $u_*$ to $u_H$. This allows to identify $u_*$ as
\begin{equation}
u_*\, =\, u_H\,  \exp\left( \frac{\mathfrak{a}}{2\,\mathfrak{b}}(\tilde{h}(u_H)-\Omega)\,\right) \simeq \, u_H\,  \exp\left(-\,\frac{\mathfrak{a}}{2\,\mathfrak{b}}\,\Omega\right)\,,
\end{equation}
where in the last step we use that $\Omega \gg \tilde h(u_H)$ which is reasonable since the $\tilde h$ variable is massive and `tries' to reach 0. Thus, in the large strain limit, $\Omega \gg 1$,  we have $u^* \rightarrow 0$. In other words, the intermediate solution \eqref{anis} extends up to very close to the boundary. Since from the UV viewpoint $u_*$ represents the crossover scale where $\tilde h$ changes from constant to logarithmic behaviour, just from dimensional analysis one expects that the subleading term in the AdS UV expansion \eqref{exp}, $\mathcal{C}_3$, must scale like $\mathcal{C}_3 \sim u_*^{-3}$. The shear strain is defined as $\varepsilon = 2\sinh (\frac{\Omega}{2})$, thus the shear stress for a large shear strain will scale with $
\nu_2^S\,=\,3\,\frac{\mathfrak{a}}{\mathfrak{b}}$ as indeed shown in fig.\ref{shearfig1}.\\

Let us return now to the solution \eqref{anis}, which implies that the  bulk metric in this limit takes a special form. Going to the coordinates $\tilde x$, $\tilde y$ where the metric is diagonal, the bulk asymptotic geometry is
\begin{equation}\label{MGLifshitz}
ds^2 \simeq \frac{1}{u^2}\left( \frac{du^2}{f_0} -  u^{-2 \mathfrak{b}^2 /\mathfrak{a}^2}  \,dt^2+  u^{2\mathfrak{b}/\mathfrak{a}} d{\tilde x}^2 +  u^{-2\mathfrak{b}/\mathfrak{a}} d{\tilde y}^2  \right)~.
\end{equation}
This geometry represents a new UV fixed point. It is attractive towards the UV, but configurations with large enough strain get close to \eqref{MGLifshitz} for a long range in $\log\,u$. Interestingly, it  exhibits both a Lifshitz dynamical exponent (between space and time directions) as well as a Lifshitz scaling with respect to $\tilde x$ and $\tilde y$ spatial directions -- that is an anisotropic scaling. 
It is quite natural that for large strain the $x$ and $y$ directions become very anisotropic. 

One might find more surprising that this occurs via a combined Lifshitz scaling both in the space and time directions. 
One way to understand this is by recalling that the models \eqref{action} can be thought of as massive gravity theories. Indeed, in the unitary gauge $\Phi^I=\delta^I_j \,x^j$ one has $X={\rm tr} \,g^{ij}$ and $Z={\rm det}\,g^{-1}$, so the scalar kinetic function  $W(X,Z)$ becomes a potential for the (spatial part of the) metric. Recall that the simplest way to obtain a Lifshitz geometry is to support it with a massive spin-1 vector field \cite{Kachru:2008yh,Taylor:2008tg,Cremonini:2014pca}. It is not so surprising, then that a massive spin-2 field can do a similar job.

Note that the existence of the anisotropic Lifshitz UV fixed point solution \eqref{MGLifshitz}
depends slightly on the choice of the potential $W(X,Z)$. 
Looking at the first of Eqns.~\eqref{simpler}, one sees that a necessary condition for $f(u)=$const. to be a solution is that $h(u)$ is logarithmic. In turn, this gives room for $W$ either to approach a constant or to vanishing towards $u=0$. In order to be consistent with the second equation, this requires that $W_h$ (which at large $h$ translates into $X W_{,X}$) is  constant along the solution. Since in these solutions $Z=u^4$, the only way that $X W_{,X}$ can reach constant is that it depends on $X,\,Z$ via the special combination $X^p\,Z^q$ with and $p\,,q$ constants which is such that every power of $Z$ can be compensated by the $X$- dependence. This is precisely what happens in our benchmark model\footnote{Notice that this choice enjoys invariance under scale transformations as a bulk quantum field theory, see \cite{Baggioli:2019elg}.} \eqref{bench}. However it also happens in much more general choices of the form  $W\big(X,Z\big)=W_0\big(X^p\,Z^q\big)$, with $W_{0}$ a free functions of one argument. (An additive function of $Z$ that vanishes at small $Z$ would lead to the same behaviour.) It is also clear that in a more generic potential that does not reduce to this form the anisotropic Lifshitz UV fixed point solution is not present. 
Still, one expects that some anisotropic solution (not completely scale invariant) should exist and dominate the response in the regime of asymptotically large strains.
\\

In any case, the (near-)Lifshitz form of the geometry is expected to impact be seen also in transport properties like the electric conductivity at finite strain (see for example \cite{Bhattacharya:2014dea}). Therefore, this suggests an avenue to possibly test whether this anisotropic Lifshitz regime occurs in real materials.

Anisotropic models with a uni-directional scalar field $\phi= \alpha z$ have been already used extensively in the literature \cite{Mateos:2011ix,Jain:2014vka,Rebhan:2011vd}. 
These models are expected to share some of the features of our solutions with large strain, as these can be represented by the scalars in the configuration $\Phi^x =\alpha \, e^{\Omega}$ and $\Phi^y = \alpha \, e^{-\Omega}$ in the limit $\Omega\to\infty$ keeping $\alpha \, e^{\Omega}$ fixed.

\subsubsection*{Elasticity bounds}

Thus far, we have discussed how to extract and understand the stress-strain curve for the black branes with solid scalar hair \eqref{stainHair}.  
One can go one step further and give an estimate on the where the stress-strain curve should terminate. As explained in \cite{Alberte:2018doe}, this can be done by studying the stability around the strained configuration. Generally speaking, this translates into a maximum strain $\epsilon_{max}$, beyond which the solutions have unstable  perturbation modes which would render them unphysical. 

The computation of the {\it elasticity bound} $\epsilon_{max}$ for the solids which can be described by EFT methods was presented recently in \cite{Alberte:2018doe}. 
We are now ready to perform the analogous computation for our holographic solids that, as emphasized above, model the special case when scale invariance is a manifest -- the solid is in a nontrivial fixed point.  

To this end, one should study the stability of the strained solutions under small perturbations. In the context of black brane solutions, this proceeds by the computation of (the dispersion relation of) the quasi-normal modes (QNMs) for the solutions \eqref{geometry}. This effort is beyond the scope of the present work, however we can already gain insight analyzing the perturbations in the {\it decoupling limit}.

More precisely, we will exploit the two following approximations. First, rather than working out the QNMs, we will look at the local propagation of fields in the bulk. 
At this level, it is much easier to identify when the propagation speed  $c_{(i)}^2$ of some mode `$i$'
develops a wrong sign,  $c_{(i)}^2<0$. By the local propagation speeds here we mean the coefficients in the $x,\,y$- gradient terms in the bulk equations of motion for the perturbations, which are functions of $u$.
Certainly, the AdS boundary conditions are such that the QNMs might turn out to be stable modes even if some speed $c_{(i)}^2<0$ somewhere in the bulk. However, one expects that the threshold where the QNMs show an instability should be near the threshold where the first mode develops $c_{(i)}^2<0$ somewhere in the bulk. In any case, using this bulk criterion is expected to place a conservative upper bound on $\epsilon_{max}$.

On the other hand, we are going to work in the {\it decoupling limit} where the mixing between scalar perturbations $\delta\Phi^I$ with the metric modes is neglected.  
In this approximation, the equations for $\delta\Phi^I$ are completely parallel to those in a $2+1$ EFT in flat space -- the  exercise that was already done in \cite{Alberte:2018doe}. On the  configurations with finite strain \eqref{stainHair} the phonon sound speeds are anisotropic, but one can still diagonalise the modes and thus obtain two characteristic speeds in the $(x,y)$ plane
\begin{equation}\label{cpm}
    c_\pm=c_\pm (\varepsilon,\varphi,u)
\end{equation}
that depend on the strain magnitude $\varepsilon$, and on the angle $\varphi$ of propagation of the sound wave with respect to one of the the principal axes (eigendirections) of the strain tensor. The subscripts $\pm$ refer to the smallest and the largest of the two speeds. (In our  holographic model, the speed in the holographic direction $u$ is always 1.)

The characteristic speeds are local, in the sense that they are defined at every slice $u=$const, but their structure in terms of the potential $W$ is identical to the one obtained in \cite{Alberte:2018doe}. Moreover, for  monomial potentials like \eqref{bench}, it is easy to see that the $u-$dependence disappears from characteristic speeds.
A conservative implementation of a stability criterium against gradient instabilities, then, is that
\begin{equation}\label{ginst}
    c_-^2>0~
\end{equation}
(for all angles $\varphi$). 
In fact, for the model \eqref{bench}, the local speeds in the $(x,y)$ plane depend on $\mathfrak{a},\, \mathfrak{b}$ in the very same way as how they depended on $A,\,B$ in the first benchmark model of \cite{Alberte:2018doe}. The condition \eqref{ginst} reads exactly the same, however the physical role of the parameters is different.

This leads to a maximum strain deformation $\epsilon_{max}$ which reduces to a certain function of $\mathfrak{a},\, \mathfrak{b}$ for the model \eqref{bench}. In the following, we discuss the obtained results, by referring to physical parameters like the scaling exponents $\nu_{1,\,2}^S$ (Eqs. \ref{nuscal1} and \ref{nuscal}) instead of $\mathfrak{a},\, \mathfrak{b}$.\\

Note that both $\nu_{1,2}^S$ scaling exponents can be bigger and lower than $1$, meaning that both \textit{softening} and \textit{stiffening} can appear\footnote{Notice how this was not possible in \cite{Baggioli:2019mck} due to the more restricted choice of potentials, corresponding in our notation to $\mathfrak{a}=\mathfrak{b}$. In this restricted class of theories, $\nu_2^S=3$ and the materials always display strain stiffening, consistent with the findings of \cite{Baggioli:2019mck}.}. Nevertheless the second scaling is always smaller than the first $\nu_2^S<2\mathfrak{a}$ because of consistency requirements \eqref{cond}.

\begin{figure}[t]
\begin{center}
\includegraphics[height=6cm]{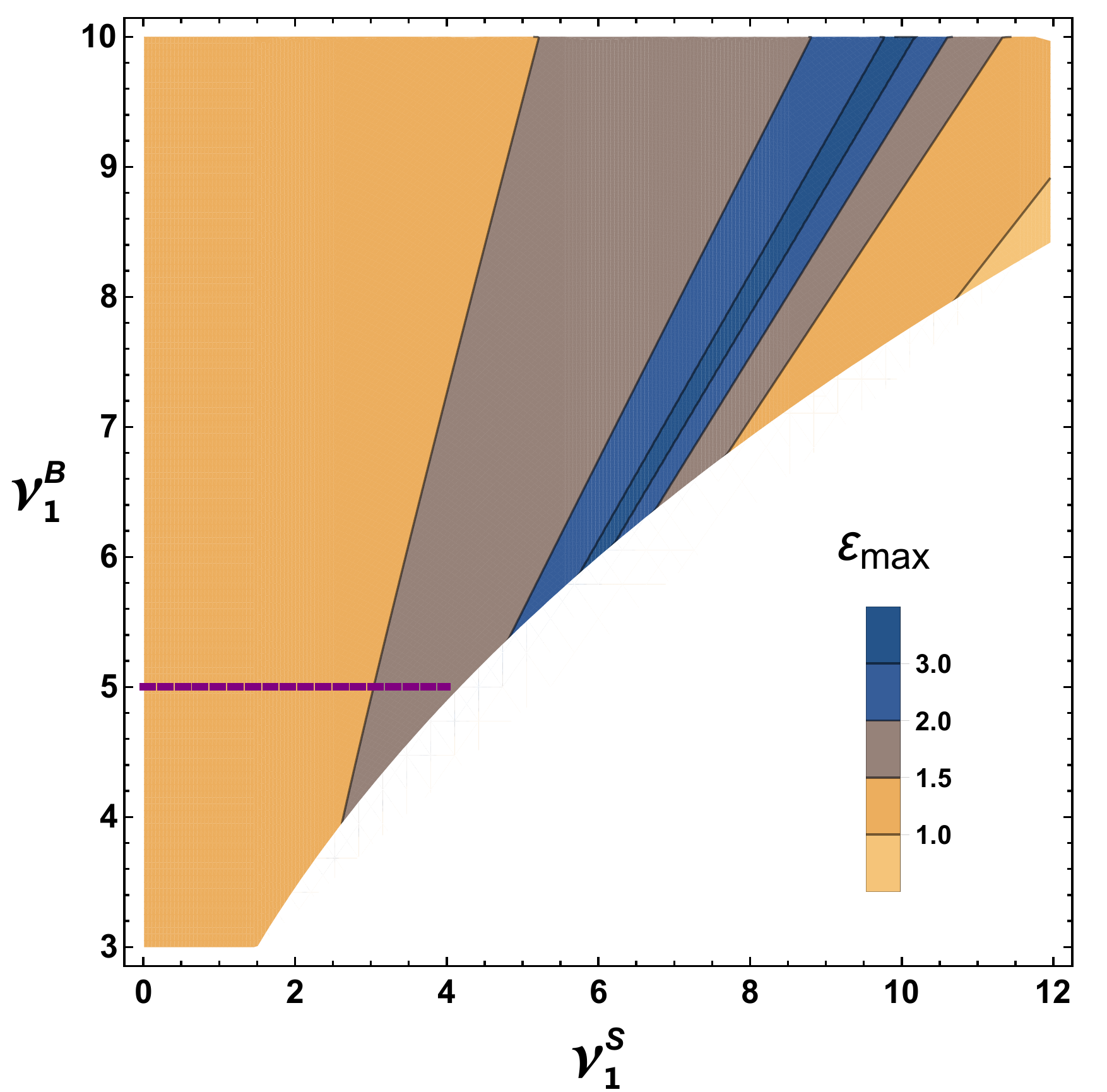}
\quad 
\includegraphics[height=6cm]{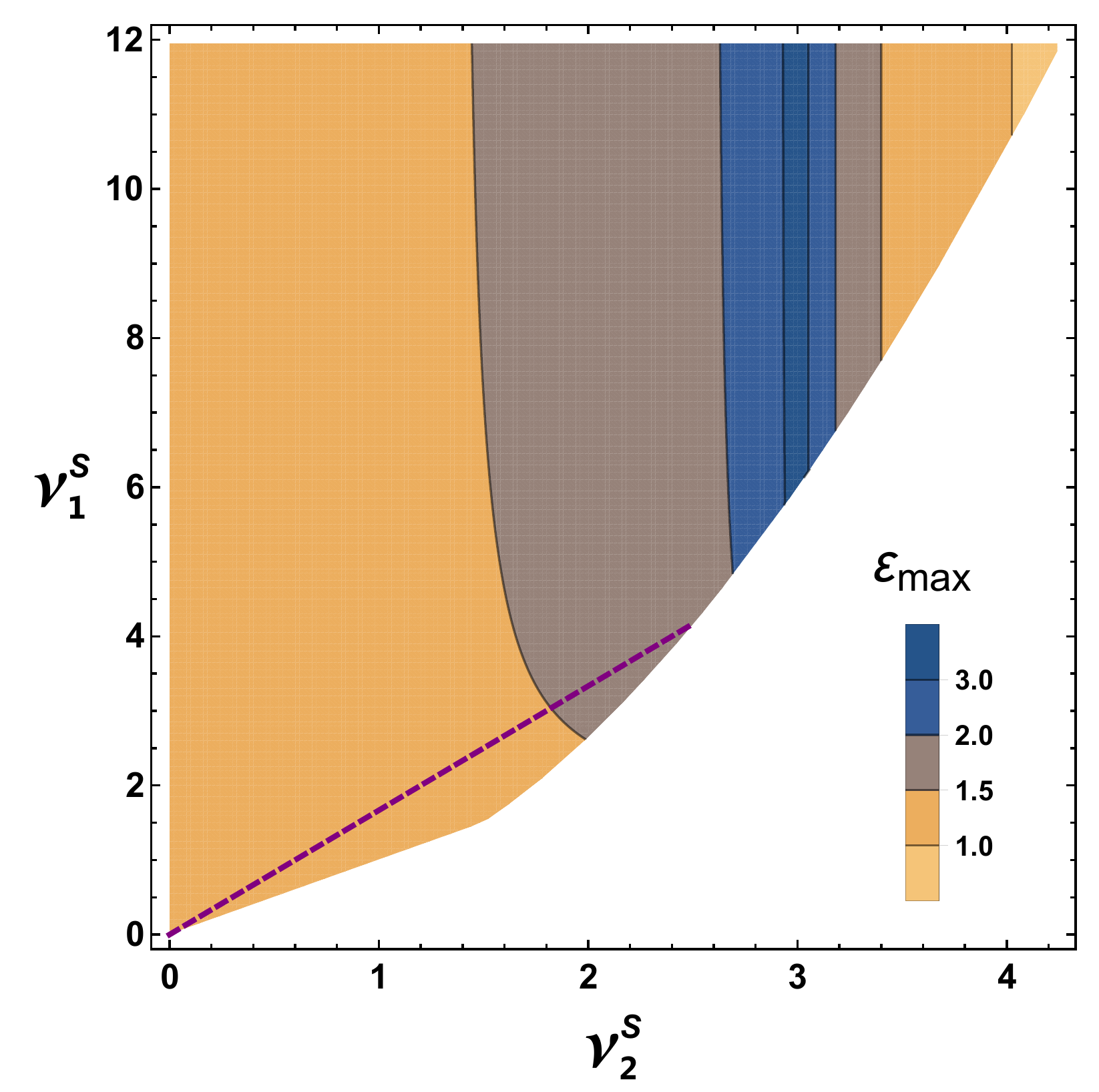}
 \caption{\textbf{Left: }Maximum shear strain, $\varepsilon_{max}$, as a function of the intermediate bulk and shear scaling, $\nu_1^B$ and $\nu_1^S$. There is one region with large $\varepsilon_{max}$ (large elasticity) around $\nu_1^S=\nu_1^B$. The limit in the strain at left and right of this region is also determined by requiring absence of gradient instability and ghosts respectively. In the white area, the longitudinal phonon is already superluminal at zero strain. 
 \textbf{Right: }
 The same plot expressed in terms of the second shear scaling exponent $\nu_2^S$. (Notice that from \eqref{nuscal1}, \eqref{nuscal} and \eqref{nuB} we can write $\nu^S_2 = 3 \; \nu^S_1/\nu^B_1$.)
The region with asymptotically large $\varepsilon_{max}$ is around $\nu_2^S=3$. 
}
 \label{CFTfig}
\end{center}
\end{figure}

Aside from the conditions shown in \eqref{cond} that restrict the values of $\mathfrak{a}$ and $\mathfrak{b}$, we also find consistency conditions on the maximum shear strain that can be applied to the system, as mentioned above. From Fig.~\ref{CFTfig}, we can already see that $\nu_2^S < \nu_1^S$, as expected. The upper bound to the maximum deformation $\epsilon_{max}$ comes from two different consistency conditions depending on the value of $\nu_2^S$: for the region $\nu_2^S>3$ we will first encounter ghosts (a change of sign in the kinetic term) at a finite value of shear strain, which will determine the value of $\varepsilon_{max}$, while in the region $\nu_2^S <3$ this value will be determined by a gradient instability (a change of sign in the speed $c_-^2$). There is a region with very large $\epsilon_{max}$ in between these two sectors, i.e. around $\nu_2^S \sim 3$, where it grows asymptotically. We can find the specific relation between $\epsilon_{max}$ and the shear scalings close to this area

\begin{equation}
\epsilon_{max} \sim \,  \left( \frac{6}{|\nu^S_2-3|}\right)^{1/4}\,.
\end{equation}

Surprisingly, the expression for the maximum deformation is the same either we are in above or below the value $\nu_2^S = 3$, and it is independent of the value of $\nu_1^S$.

The subluminality constraint is also included Fig.~\ref{CFTfig}, in the following way. The local  speeds in the bulk \eqref{cpm} are allowed to be possibly superluminal, as the speeds in \eqref{cpm} do not give directly any physical phonon speeds. A proper limit would require the computation of the QNM dispersion relations at finite strain, which is outside the scope of this work. 
As a first step, though, we include the known results for the QNMs at vanishing strain \cite{Ammon:2019apj,Baggioli:2019abx}, which certainly places a bound and one expects it gives an idea of the kind of bound one should obtain. We thus cut out the white region, which is where  the physical longitudinal phonon QNM would be superluminal at $\varepsilon=0$.
By continuity, the true subluminality bound should make $\varepsilon_{max}$ vanish smoothly close to the edge of the white areas in Fig. 4. This has the effect to decrease $\varepsilon_{max}$ near the white edge and it should also render the actual plot less blue in the region $\nu_1^B\sim\nu_1^B\sim 6$, for instance. However, given the different slopes of the `blue ray' and of the edge of the white area, one expects that the high elasticity region (the blue ray) persists mostly.

\subsubsection*{Bulk deformations}
\begin{figure}[t]
\begin{center}
\includegraphics[height=7cm]{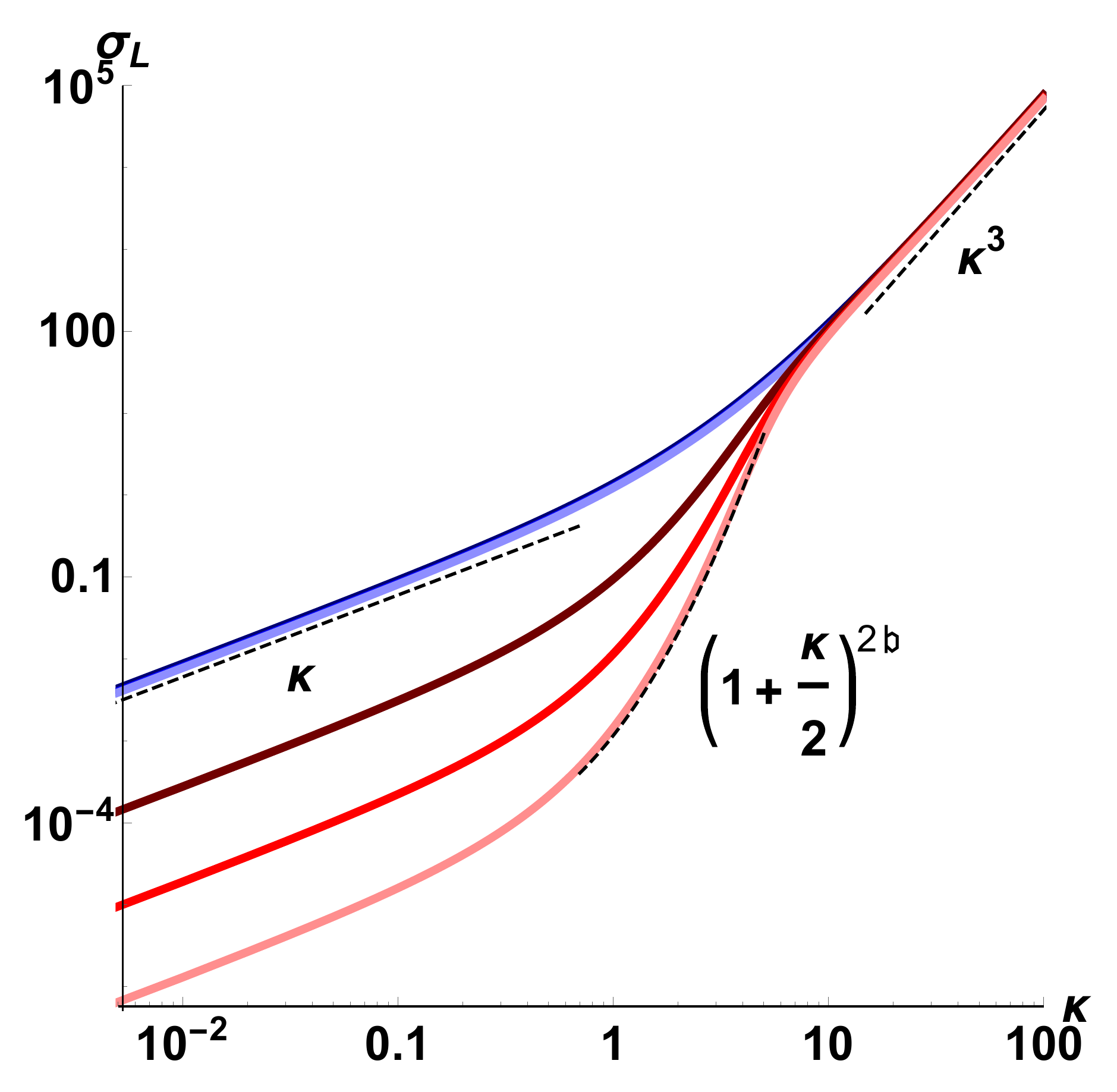} \quad \includegraphics[height=7cm]{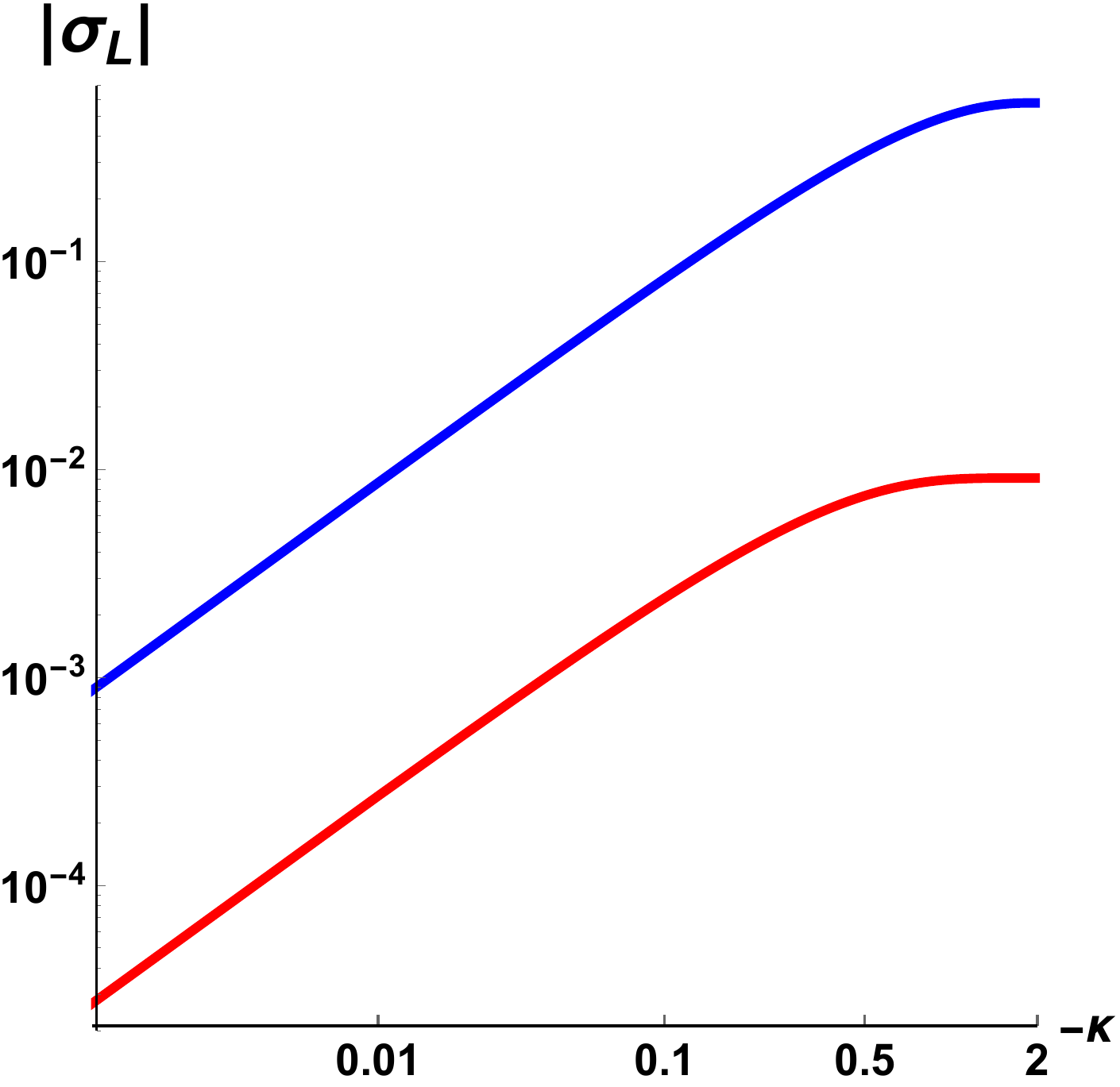}
 \caption{\textbf{Left: }Non linear bulk elastic response for various potentials $W(X,Z)$ at small temperature $T/m=0.01$ (blue lines) and $T/m=1$ (red lines). At large deformations all the lines asymptote the universal scaling \eqref{univ} $\sim \kappa^3$. At large temperatures there is a transition between linear and cubic scaling that grows as $\alpha^{2\mathfrak{b}}\,=\,\left(1+\frac{\kappa}{2}\right)^{2\mathfrak{b}}$. \textbf{Right: }Absolute value of $\sigma_L (\kappa)$ for $\mathfrak{a}=\mathfrak{b}=3$ and $T/m = 0 , 1$ (blue, red) as a function of $-\, \kappa$.}
 \label{figbulk}
\end{center}
\end{figure}

We can as well consider the bulk response beyond the linear approximation. In this section, we will neglect the finite temperature corrections; as a consequence the results presented here are robust at small temperatures $T/m \ll 1$ but they will probably get finite $T$ corrections elsewhere.
In order to compute the non linear bulk response and avoid mixing the different deformations we set the shear strain $\varepsilon=0$. In that case, the equilibrium configuration \eqref{scal} is set by $\Phi^I=x^I$; this means that a bulk deformation $\kappa=\partial \cdot \phi$ corresponds to $\kappa=2(\alpha-1)$. Now, in analogy with linear response, we can compute the response in the pressure $T_{xx}=T_{tt}/2$, \textit{i.e.} the longitudinal stress, with respect to the longitudinal strain in a full non linear form. We define the bulk stress as $\sigma_L=T_{xx}(\kappa)-T^{eq}_{xx}$. In more details for our model \eqref{bench} we obtain:
\begin{equation}
\sigma_L(\kappa)\,=\,\frac{m^2\, \left(\left(\frac{\kappa}{2}\, +\,1\right)^{2\, \mathfrak{b}}\, u_{h,\kappa}^{2\,\mathfrak{b}\,-\,3}\,-\,u_{h,\kappa = 0}^{2\,\mathfrak{b}\,-\,3}\right)
   }{2\, (2\, \mathfrak{b}\,-\,3)}\,+\,\frac{1}{2}\left(\frac{1}{u_{h,\kappa}^3}\,-\,\frac{1}{u_{h,\kappa = 0}^3}\right)\,,
   \label{bulkdef}
\end{equation}
where with $u_{h,\kappa = 0}$ we mean the value of the BH horizon in absence of any bulk strain, \textit{i.e.} $\alpha=1$. The sign of $\kappa \equiv \partial \cdot \vec{\pi}$ can be positive, \textit{i.e.} a compression, or negative, \textit{i.e.} an expansion.\\
The results for various potentials are shown in fig.\ref{figbulk}.
At large bulk strain $\kappa \gg 1$, we find a universal scaling
\begin{equation}
\sigma_L\,\propto\,\kappa^3\,,\label{univ}
\end{equation}
which is a consequence of conformal invariance and it can be generalized to $\kappa^D$, with $D$ the number of spacetime boundary dimensions. The scaling can be immediately obtained analytically just realizing that in the limit $\kappa \gg 1$ the radius of the horizon scales like $u_H \propto 1/\kappa$ \footnote{This scaling comes from exactly the identical arguments given for the so-called \textit{incoherent limit} in a slightly different context \cite{Baggioli:2016pia,Baggioli:2017ojd}.}. 
Furthermore, this result is in perfect agreement with the EFT computations \cite{Alberte:2018doe}.
Similarly to what happens in shear deformation, at large temperatures, there is an in between scaling that goes as $\alpha^{2\mathfrak{b}}\,=\,\left(1+\frac{\kappa}{2}\right)^{2\mathfrak{b}}$, as can be seen in fig.\ref{figbulk}. Notice that there are again two types of deformation:  isothermal ($T=const.$) and adiabatic ($s=const.$). In this last case, the scaling is always $\sigma_L\,\propto\,\kappa^3$, while the intermediate scaling only shows up for isothermal deformations.

Therefore, we can define two different bulk scalings, one that appears at intermediate strains and finite temperature and another that appears at very large strain or low temperatures, just like what happens for the shear deformation. These two bulk scalings are 
\begin{equation}\label{nuB}
    \nu_1^B \equiv 2\,\mathfrak{b} \quad , \quad  \nu_2^B \equiv 3\,,
\end{equation}
where the first scaling is always equal or larger than the second: $\nu_1^B\geq \nu_2^B$.\\

So far, we have focused on bulk deformations with $\kappa > 0$ but we have not discussed negative values of $\kappa$, \textit{i.e.} expansion. In this case, in order to produce a full expansion we need to go to the limit where $\alpha\rightarrow 0$ which corresponds to $\kappa = -2$. We can see from \eqref{bulkdef} that for low temperatures the stress goes as $\sigma_L \sim (\alpha^3-1) = ((\frac{\kappa}{2}+1)^3-1)$ and for temperatures high enough the scaling would be $2 \mathfrak{b}$ instead of cubic. However, we do not get to see these non-linear scalings due to the short range of values $\alpha$ and $\kappa$ have during the expansion. A couple of examples at different temperatures are shown in the right panel of Fig.~\ref{figbulk}, where we just see a linear scaling that gets saturated at some finite value. There, we also find that the amount of stress needed to produce a full compression is finite. We are not aware of any experimental signature of scale invariance in the linear and non-linear elastic response. Our results suggest that scale invariance should play an important role, providing universal scaling which can be possible tested at quantum critical points or within the quantum critical region.

\section{Solids with spontaneously broken scale invariance}
\label{section4}

In this section we present the analysis of the nonlinear response, for models of solids that realize scale invariance (SI) as a spontaneously broken symmetry. 
This case can be treated using EFT methods in \cite{Alberte:2018doe}.
Linear elasticity of this systems has been recently discussed in \cite{Baggioli:2019elg}. The goal now is to extend the analysis to the nonlinear regime properties and to compare it with the {\it manifest} SI case studied in the previous section with AdS/CFT techniques.

\subsection{Nonlinear response from EFT methods}
In order to study the nonlinear response of a solid with spontaneously broken scale invariance we are going to employ EFT methods. We have already presented an EFT of a solid in \eqref{action}, which has been studied in the past in \cite{Dubovsky:2011sj,Nicolis:2013lma,Nicolis:2015sra,Alberte:2018doe}, but now we want to consider only Lagrangians that are invariant under scale transformations. Therefore, we are going to demand that the Lagrangian that we introduced in \eqref{action} is invariant under 
\begin{equation}
x^\mu \rightarrow \lambda^{-1}\,x^\mu , \quad \phi^I \rightarrow \lambda^\Delta \phi^I \,,
\end{equation}
with some `weight' $\Delta$ for the fields $\phi^I$ that will depend on the potential, as can be seen in \cite{Baggioli:2019elg}. We find that the most general potential that we can use must be of the form of
\begin{equation}
V_{SI}(X,Z)\,=\,Z^{\frac{1+\omega}{2}}\,f\left(x\right)\,,
\label{benchmark}
\end{equation}
with $x \equiv X / \sqrt{Z}$ and 
$$
\omega=\frac{1-(d-1)\,\Delta}{(d-1)(\Delta+1)}\,,
$$ 
which is also identified as the equation of state parameter, $\omega = p/\rho$.  Notice that, although the potentials \eqref{benchmark} and \eqref{model} look alike, their relation is not trivial as the framework where they are used is completely different. The comparison between the two should be done through physical properties of the system they describe such as the non-linear scalings.  Moreover, this potential is not describing a system with conformal invariance unless we take the particular case where $\omega = 1/2$.\\

\subsection{General results}

The non-linear bulk scaling is determined by $\omega$ and is completely independent of the function $f(x)$. This scaling is given by 
\begin{equation}
    \nu_B = 2\,(1+\omega )~.
\end{equation} 
Moreover, the bulk strain $\kappa$ is unconstrained, as opposed to the shear strain $\varepsilon$. 
Looking at Fig.~\ref{monoEFT}, we can see that the case of a monomial potential the scaling is restricted to satisfy $2< \nu_B < 4$. For a more general $f(x)$, we can find that the limit is still the same, \textit{i.e.} we cannot have $\omega > 1$ without gradient instabilities or superluminal modes. Thus, neither the non-linear bulk scaling nor the constraints of it depend on the shape of $f(x)$.

\subsection{Benchmark models}

The results related to the shear response are potential dependent, so we will need to specify what type of potential we want to study. In particular, we will take the two cases considered in \cite{Alberte:2018doe}, which are:
\begin{enumerate}
    \item The first possibility we will study is the case where $f(x)$ is a monomial, \textit{i.e.} 
    \begin{equation}
        \label{EFTb1}
    f(x) = x^{\nu_S / 2}
    \end{equation}
    with $\omega$ as a free parameter. This simple potential realizes a power-law scaling in the stress for large deformations. This potentials will display a power-law behaviour in the stress-strain relation for the shear channel, which will be determined by $\nu_S$, i.e. we will have that at large deformations $\sigma_S \sim \epsilon^{\nu_S}$. 
    
    \item The second possibility we are going to consider is taking both 
    \begin{equation}
        \label{EFTb2}
    f(x) = 1 + v^2 x^{\nu_S/2}
    \end{equation}
    and the limit $\omega\rightarrow 0$. The advantage of this potential is that the speeds of the phonon modes will be realistic (i.e. much smaller than the speed of light) as long as $v^2 \ll 1$, whereas the monomial potential has relativistic modes for generic values of $n$ and $\omega$. The shear response will only come from the $x$-dependent term, so the power-scaling of the shear response will be determined by $\nu_S$.
   
\end{enumerate}

\begin{figure}[t]
    \centering
    \includegraphics[height=6.5cm]{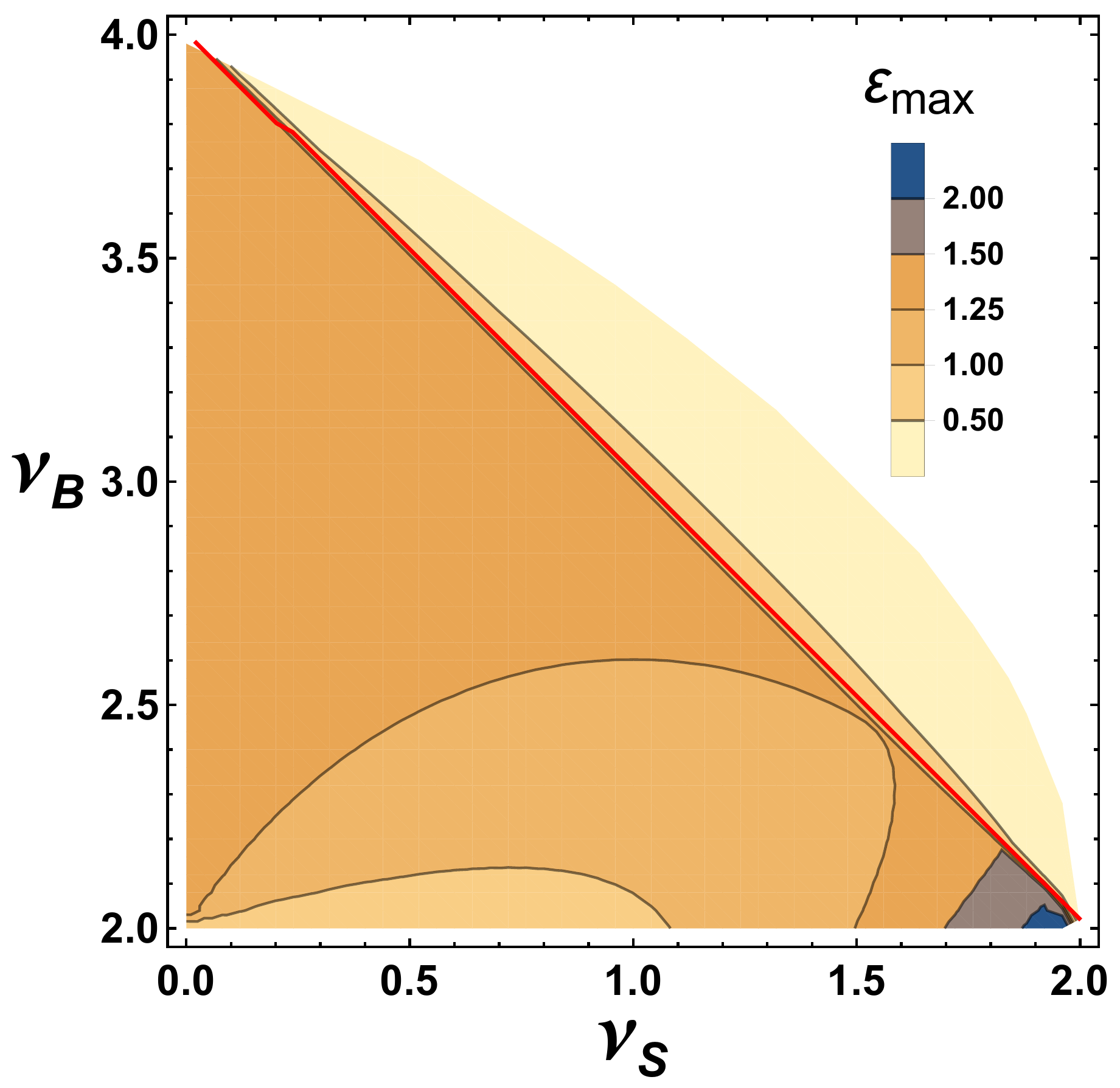}
    \quad 
    \includegraphics[height=6.65cm]{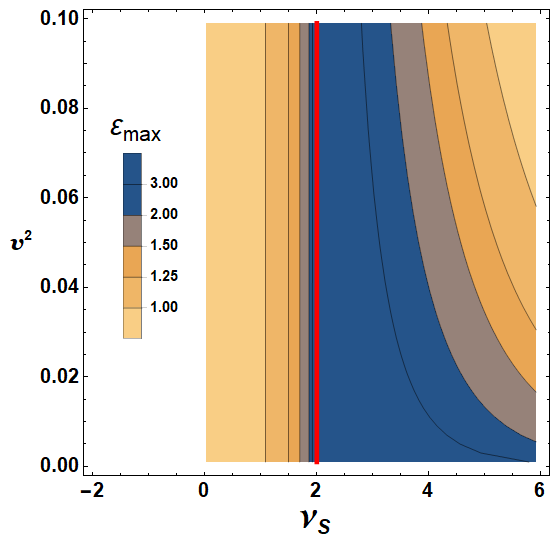}
    \caption{\textbf{Left: }The maximum allowed strain $\varepsilon_{max}$ within the allowed region for the monomial EFT case $f(x)=x^{\nu_S/2}$. The left, bottom and right edges are respectively given by: gradient instability, positivity of the bulk modulus and subluminality. The red line separates the region where $\varepsilon_{max}$ is controlled by gradient instability and subluminality. \textbf{Right: }The maximum allowed strain $\varepsilon_{max}$ within the allowed region.
 SSB case in the particular case where $\omega = 0$ and $f(x) = 1+v^2\,x^{\nu_S/2}$. The left and bottom edges are respectively given by: gradient instability and positivity of the bulk modulus.
The maximum shear strain is determined by gradient instability for $\nu_S<2$, subluminality for $\nu_S>2$ and unconstrained for $\nu_S=2$.}
    \label{monoEFT}
\end{figure}

The discussion about the  shear scaling $\nu_S$ of the spontaneous broken scale invariance (SBSI) case is of course sensitive to the form of the function $f(x)$. The simplest non-trivial example of potential one can think of is a monomial, ie. $f(x)=x^{\nu_S/2}$, which is shown in Fig.\ref{monoEFT}. In this particular scenario both the bulk and shear scalings are constrained, in particular we find that $2< \nu_B < 4$ and $0 < \nu_S < 2$. As with the solids with manifest scale invariance, here we can find one region where the maximum strain sustained by the material is significantly large. The most elastic region is found close to $\nu_S \sim 2$ and we find that for this scaling $\varepsilon_{max}$ reads as
\begin{equation}
    \varepsilon_{max} \simeq \sqrt{2} \,\left(\frac{1}{2-\nu_S}\right)^{1/4} \quad \mbox{for} \quad \nu_S\lesssim 2\,.
\end{equation}
The problem with this kind of potentials is that the speeds of the phonons are excessively large. Typically one would expect that the speeds of these modes are not bigger than $\sim 10^{-4}$ times the light speed in `earthly' materials (needless to say, this concern does not affect relativistic solids such as  neutron star interiors). This constrains very much the possible scalings one can realize in a realistic scenario, specifically we would be restricted to $\nu_S$ and $\nu_B-2$ not bigger than $10^{-8}$.\\

If we do not restrict ourselves to the most simple case, we can describe a solid with slower phonon modes. For this, we are going to take $\omega \rightarrow 0$ and $f(x) = 1 + v^2 \, x^{\nu_S/2}$ with $v^2\ll 1$ as proposed in \cite{Alberte:2018doe}.
 This form of potential ensures that the speed of the phonon modes are small at least for small shear strain. The shear scaling is forced to be positive $\nu_S >0$ and the maximum shear scaling is 
\begin{equation}
\nu_S^{max} = \frac{2\,(1+v^2)}{v^2},
\label{numaxeft}
\end{equation}
thus $\nu_S^{max} \gg 1$ for $v^2 \ll 1$. The region with $\epsilon_{max}$ large is found around $\nu_S \lesssim 2$ and also in the region where $\nu_S > 2$ and $v^2\ll 1$, specifically 
\begin{equation}
    \varepsilon_{max} \simeq \,\sqrt{2} \,\left(\frac{1}{2-\nu_S}\right)^{1/4}  \quad \mbox{for} \quad \nu_S\lesssim 2\,,
\end{equation}
\begin{equation}
\varepsilon_{max} \simeq \, \sqrt{2}\,\left(\frac{2}{v^2 \, (\nu_S-2)}\right)^{\frac{1}{\nu_S}} \quad \mbox{for} \quad \nu_S \gtrsim 2\,,
\end{equation}
where we have taken the limit $\omega \rightarrow 0$ and the last $\epsilon_{max}$ is valid for $\nu_S>2$ not necessarily close to $\nu_S = 2$  as long as $v^2\ll 1$, as can be seen in Fig.\ref{monoEFT}.

\begin{figure}[htpb]
\begin{center}
\includegraphics[height=7cm]{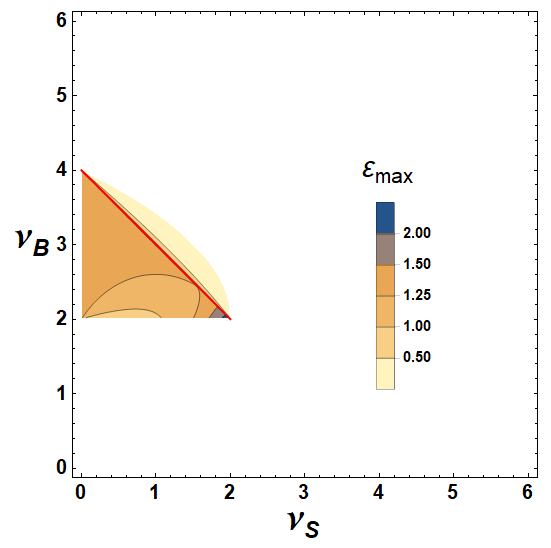}\quad \includegraphics[height=7cm]{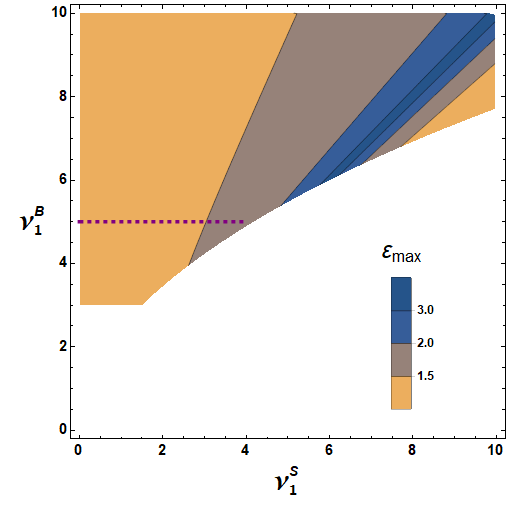}
 \caption{Summary of the main differences between the nonlinear elasticity bounds found for materials with spontaneously broken (left) or manifest (right) scale invariance. Both materials present power-law stress-strain relations $\sigma\sim\varepsilon^\nu$ (at $\varepsilon\gg1$) with exponents $\nu^B$  and $\nu^S$ for pure-bulk and pure-shear deformations. Both the values of the maximum allowed deformation $\varepsilon_{max}$ as well as the range of values of the allowed exponents differ substantially.}
 \label{comparison}
\end{center}
\end{figure}

\section{Comparison}
Let us now compare the nonlinear response obtained in the manifest/spontaneously broken SI cases. 
In order to make the comparison as  meaningful as possible, we will fix the physical observable -- the  stress-strain curve to be of the type $\sigma\sim \epsilon^\nu$  for both bulk and shear deformations with exponents $\nu_{S,\,B}$ -- and compare  the elasticity bounds in the manifest or spontaneously broken case. 
Our  holographic examples present two power-law regimes, but we will restrict attention to the first (`intermediate') one labeled with exponents $\nu_1^{S,\,B}$ because the second one is only realized at extremely large deformations (at finite temperature). \footnote{Another advantage of focusing on the intermediate scaling is that they allow for a continuum range of bulk exponents $\nu_1^{B}$ whereas the asymptotic one is fixed by conformality to $\nu_2^{B}=3$.}

The results shown in Sections \ref{section:setup} and \ref{section4} make manifest that the universal elasticity bounds that can be obtained from low-energy effective methods differ significantly depending on how scale invariance is realized (as a manifest or a spontaneously broken symmetry).
The main differences are basically summarized in Fig.~\ref{comparison}, where we repeat here the relevant plots in Figs.\ref{CFTfig} and \ref{monoEFT} but in a comparable scale, for the sake of comparison. 

\begin{comment}
\OP{
Let us remark that one might at first expect that the  elasticity bounds in the manifest SI could be obtained exploiting the EFT methods as shown in \cite{Alberte:2018doe}. 

Using Eqs. \eqref{integral} and \eqref{sigma} we see that the {\it energy function} ${\cal E}^{{}^{CFT}}$ (defined on constant strain configurations) for the manifest SI case obeys
\begin{equation}
{\cal E}^{{}^{CFT}}_{\tilde X}(\tilde X,1) \equiv \frac{m^2}{2}\,
 \int_0^{u_H}\,
 W_X\left(\frac{\tilde X\,\zeta^2}{2},\zeta^4 \right)\,
\frac{d\zeta}{\zeta^2} \;+\; {\cal O}(m^4)
\label{VCFT}
\end{equation}
with $u_H$ possibly being a function of  $\varepsilon$ or not depending on whether one works at fixed $T$ or entropy density (i.e. fixed $u_H$). This identification follows once one recognizes $\tilde X\equiv 2+\varepsilon^2$.
%
This equation can in turn be integrated once in $\tilde X$ to obtain an expression for the energy function ${\cal E}^{{}^{CFT}}$ in terms of $W$.

Let us insist that this `equivalence' is only meaningful to mimic the stress-strain curves, but no more. 
The two theories are completely different and so the way how the stress-strain curve (or the energy function) affects other nonlinear quantities (like $\varepsilon_{max}$) differs substantially.
In particular, we expect that the elasticity bounds that could possibly be extracted by treating the CFT as an EFT with potential \eqref{VCFT} to be wrong. The usefulness of the identification \eqref{VCFT}, then, looks very limited. 
}

\end{comment}

It is clear from Fig.~\ref{comparison}, the elasticity bounds present substantial differences, both in the range of allowed values in the $\nu_S -  \nu_B$ plane where finite deformations can be reached ($\varepsilon_{max}\sim1$), as well as in the region where the deformations can be large ($\varepsilon_{max}\gg1$).

In fact, if we restrict the CFT analysis to the models that exhibit massless phonons (which requires $\nu_1^B>5$) then the areas in the $\nu_S -  \nu_B$ plane are almost disconnected, with the spontaneously broken case covering lower values of $\nu_B$ -- globally in the  $2<\nu_B<4$ window. 

Similarly, the shear exponent is constrained to $0<\nu_S<2$ for the SB case while it is basically unbounded in the manifest case. 
Another major difference is the presence in the manifest case of a `very elastic band' (the bluish area)  in the region near \footnote{Interestingly, real-world rubbers are well fitted by power-law stress-strain curves with exponents satisfying \eqref{blueray} \cite{enlighten70333,Alberte:2018doe}.
}
\begin{equation}
    \label{blueray}
\nu_1^S = \nu_1^B~.
\end{equation}
This is the region that allows for {\it black rubber} - like holographic duals. It is worth to emphasize two properties about these solutions: i) the band in parameter space start at $\nu_{1}^{B,\,S} \gtrsim 6$ which corresponds to a rather stiff behaviour; ii) these values of  $\nu_{1}^{B,\,S}$ are far from the free scalar limit ($\mathfrak{a}=\mathfrak{b}=1$, corresponding to $\nu_{1}^{B,\,S}=2$). Thus, black rubbers require the presence of scalars with non-canonical (non-linear) kinetic terms in the bulk.  

For the EFTs of Sec.~4 instead, the {\it hyper-elastic} %
region collapses down from a strip to basically a `dot'-like area. Let us remark that this last feature happens  in the first EFT  benchmark \eqref{EFTb1} but not in the second one \eqref{EFTb2} (which displays small sounds speeds and also an {\it elastic band} at $\nu_1^S \sim \nu_1^B$). However, we prefer to keep the comparison at this level because the benchmark  \eqref{EFTb2} depends on an extra parameter.\\

As mentioned above, the reason to introduce the second benchmark model for in Sec.~4 is to be able to have realistic phonon  speeds -- much smaller than the speed of light $c$. A fair question, then, is whether this is also possible in the AdS/CFT framework or we are forced to have phonon speeds of the order of $c$. This issue has been already addressed in \cite{Baggioli:2019elg}, where it is claimed that a solid with manifest SI and small phonon speeds is achievable. 
In the holographic set-up the speeds of the phonons are obtained by finding the spectrum of the quasi-normal modes, which have the form 
\begin{equation}\label{QNM}
\omega = c_s k - i\,D\,k^2\,\dots
\end{equation}
for both transverse and longitudinal modes, as checked in \cite{Alberte:2017oqx,Ammon:2019apj}. 
The low energy dynamics of these solids with manifest SI are described by a low energy CFT (and thus an IR fixed point). It is conceivable that the Lorentz group that emerges in the IR has a different light-cone speed $c_e$, and thus the space-time metric is $ds_e^2 = -c_e^2\, dt^2 + dx^i dx_i$. 
This has a crucial impact in \eqref{QNM} if we compare the results between a Lorentz invariant theory with a light-cone speed $c$ and one with $c_e$, as we will obtain a rescaling in the dispersion relation as $\omega \rightarrow (c/c_e) \,\omega$. Therefore, for $c_e \ll c$ the phonon speeds would get suppressed as
\begin{equation}
c_s \rightarrow \frac{c_e}{c}\, c_s\,\ll c\,.
\end{equation}
If this is the case, we could also have a ``realistic'' solid with smaller speeds and, in addition, the
white areas --limited by superluminal constraints-- in Figure \ref{CFTfig} would be enlarged.

\section{Conclusions}\label{section:conclusions}

We have analyzed the nonlinear elastic response in materials with scale invariance from the low-energy perspective, using effective field theory and holographic methods. 
The advantage in these effective methods is that they are mainly based on how symmetries are realized and therefore they can help to understand the nonlinear behaviour `universally’, that is,  independently of the microscopic details of the material. 
The constraints imposed by how symmetries are realized imply  nontrivial relations among different low-energy observables, especially the ones that encode  the non-linear structure of the theory — the interactions in the material.

In order to illustrate the appearance of these constraints from the low energy theories (and their dependence on how symmetries are realized), we have focused on the example provided by the {\it elasticity bounds}:  the maximum deformability that a material may  withstand in a reversible form. These were discussed in \cite{Alberte:2018doe} in the case for materials with no manifest scale invariance. In this work, we have obtained these bounds for scale invariant  (SI) materials, which can be of two types: with manifest SI or with spontaneously broken SI. The latter case can be described using the EFT methods and so is a particular case of those discussed in \cite{Alberte:2018doe}. For the manifest SI case, instead, we have used holographic models.
In order to include the two key ingredients (elasticity and manifest scale invariance), we have studied the simple holographic models of ‘massive gravity’ type. These are well-defined effective field theories in  (asymptotically) AdS spacetime and they allow for a straightforward interpretation as a scale invariant field theories. 

The main result of this work is to show how the elastic response can be extended to the full non-linear regime by obtaining the full stress-strain curves. The procedure is straightforward, and it gives rise to a very rich phenomenology of nonlinear elasticity behaviours which can easily be extended to other holographic models. 

To our knowledge, this is the first time that the full non-linear elastic response of AdS black brane geometries is presented by extracting the corresponding stress-strain curve. 
Previous studies have discussed the elastic response only in the linear approximation. Ref. \cite{Baggioli:2019mck} discussed the visco-elastic oscillatory (that is, non-static) response and \cite{Biasi:2019eap} an out-of-equilibrium similar setup.  \\

As in \cite{Alberte:2018doe}, in order to make progress, we have assumed models that, by assumption, display power-law stress-strain curves, $\sigma(\varepsilon)\sim \varepsilon^\nu$ with  some constant exponent $\nu$ at large strains $\varepsilon\gg1$. We have constructed the stress-strain curve and also followed how the appearance of pathologies (gradient instability, ghosts, or superluminal propagation) appears as a function of $\varepsilon$.
We have then obtained  how the maximum strain $\varepsilon_{max}$ depends on the exponents $\nu^S$ and $\nu^B$ (for shear and bulk transformations respectively). 

We highlight three aspects of our obtained results. First, we find that the elasticity bounds for the solids with manifest SI (the holographic models) differ substantially from the ones for the solids with spontaneously broken SI (the EFT models), as shown clearly in Fig.~\ref{comparison}. Both the ranges of allowed exponents and the values of the deformability $\varepsilon_{max}$ disagree by $\mathcal{O}(1)$ factors. 
Our interpretation of this discrepancy is that it is physical and due to the fact that SI is realized differently in the two cases, implying that the nonlinear constraints in the theories must differ.   

The second aspect to illustrate is that our holographic models present the interesting peculiarity that they exhibit actually two different regimes of power-law stress-strain curves, $\sigma(\varepsilon)\sim \varepsilon^\nu$,  with different exponents at moderately large $\varepsilon$ (say, $1<\varepsilon<10$) and asymptotically large $\varepsilon$. This is due to the fact that these models contain a UV anisotropic Lifshitz fixed point. It is unclear whether this feature is only specific to the present model or whether it should be more generic in (very) elastic solids with manifest SI. Neither it is clear how relevant this feature is for realistic materials as this second scaling regime only appears for extremely large deformations, $\varepsilon \gtrsim 10$, before which the solutions already show some instability. 

Finally, we stress that the holographic models that exhibit highest elasticity share features surprisingly similar to familiar real-world elastomers. Indeed, the models that allow black brane solutions which are stable under largest deformations (largest $\varepsilon_{max}$) turn out to have power-law exponents $\nu$ in the stress-strain curves $\sigma\sim \varepsilon^\nu$ which are: i) of order a few; ii)  similar for pure-shear and pure-bulk $\nu_S=\nu_B$ (corresponding to the blue stripe in Fig 7.b). Intriguingly, this is what happens for natural rubber and other elastomers \cite{enlighten70333} -- and it motivates us to call these solutions {\it black rubber}.\\

Very importantly we would like to comment on the possible applications of this framework to realistic systems.\\

(I) The electric transport properties of quantum critical materials have been subject of lot of recent efforts especially in connection to the ``anomalous'' scalings found in Strange metals. More recently the question whether
also phonons and elastic properties can display surprising and interesting features in quantum critical materials has emerged. More specifically there are preliminary indications that phonons in quantum critical systems can exhibit glassy or viscoelastic
features \cite{ishii2019glass}. Moreover the role of these viscoelastic properties has been discussed in connection to the possible implications on the onset of (high-Tc) superconductivity \cite{setty2019glass}.
The framework just presented can provide a useful methods and possible observable
predictions in this directions. The utility of these holographic models has been already proven in understanding the glassy features of amorphous solids and ordered crystals in \cite{Baggioli:2018qwu,baggioli2018soft,baggioli2019unified}.\\

(II) Let us also mention the growing interest related to the non-linear mechanical characterization of critical materials (e.g. High-$T_c$ superconductors) due to their technological applications  \cite{Scheuerlein:2017kjv,Scheuerlein:2017zka}. Given the absence of robust computational methods, one may not rule out that the holographic methods such as presented here may provide useful insights.\\

(III) Following \cite{Baggioli:2019mck}, a full non-linear characterization of the mechanical response of the holographic homogeneous models \cite{Baggioli:2014roa} considered in this work could definitely shed light on their physical nature, which is still open \cite{Ammon:2020xyv,Baggioli:2020nay}.\\

Finally, we remark  a perhaps more technical point that we think is interesting from the gravitational point of view. The computation of the mechanical response to non-linear shear deformations translates in the AdS/CFT dictionary into finding exact black hole solutions with finite shear deformation (and shear stress) in the transverse directions. This implies that these solutions have a non-trivial, nonlinear spin-2  mode, and we devoted some effort to derive them in Sec.~3. These solutions are similar to the nonlinear gravitational waves and `pp-wave' solutions in General Relativity in that they also have a nonlinear spin-2 (transverse-traceless) mode 
However, in our models the solutions  are static, which is a direct manifestation of the {\it massive} character of the metric in these models. 
Moreover, the spin-2 mode `sticks out' of a black brane horizon. In this sense, then, these solutions can be thought of as branes with {\it spin-2 hair}.  We are unaware of solutions of this kind in the literature, but we find it remarkable that they are exist, and are tied to the nonlinear elastic response.

\section*{Acknowledgements}
We acknowledge and thank Sébastien Renaux-Petel and Ke Yang for collaboration at the early stages of this project. We thank Martin Ammon, Tomas Andrade, Alex Buchel, Amadeo Jimenez-Alba, Niels Obers, Napat Poovuttikul, Kostya Trachenko and Alessio Zaccone for useful discussions and comments about this work and the topics considered. 
MB acknowledges the support of the Spanish MINECO’s “Centro de Excelencia Severo Ochoa” Programme under grant SEV-2012-0249. OP acknowledges support by the Spanish Ministry MEC under grant FPA2014-55613-P and the Severo Ochoa excellence program of MINECO (grant SO-2012-0234, SEV-2016-
0588), as well as by the Generalitat de Catalunya under grant 2014-SGR-1450. 

\appendix

\section{Holographic Stress Tensor}\label{appendix:stress}
We can explicitely compute the boundary stress tensor $T^{\mu\nu}$ following \cite{Balasubramanian:1999re}. In particular, for asymptotic AdS$_4$ spacetime, we have:
\begin{equation}
T^{\mu\nu}\,=\,\frac{1}{8\,\pi\,G}\,\left(\Theta^{\mu\nu}\,-\,\Theta\,\Xi^{\mu\nu}\,-\,2\,\Xi^{\mu\nu}\,-\,G^{\mu\nu}_{\Xi}\right)\,,
\end{equation}
where we set the AdS length $l=1$. The metric $\Xi_{\mu\nu}$ is the boundary metric and $\Theta^{\mu\nu}$ the extrinsic curvature of the boundary surface, with $\Theta$ its trace. Our boundary metric is given by 
\begin{equation}
\Xi_{ij}= \frac{1}{u^2} \begin{pmatrix} e^{-\chi(u)} \, f(u) & 0 & 0 \\
0 & \cosh h(u) & \sinh h(u) \\
0 & \sinh h(u) & \cosh h(u)
\end{pmatrix}
\end{equation}
and it is clearly flat, implying $G^{\mu\nu}_\Xi=0$. We can define the normal vector to the boundary as:
\begin{equation}
n_\mu \, = \, \left(0\, ,\, 0\, ,\, 0 \, ,\,\frac{1}{\sqrt{g^{uu}}}\right) ,
\end{equation}
where $g^{uu}=u^2 f(u)$. The extrinsic curvature can be defined as usual
\begin{equation}
\theta _{\mu \nu} \, = \, \frac{1}{2} \left( \nabla_\mu n_\nu \, + \, \nabla_\nu n_\mu \right), 
\end{equation}
and it reads 
\begin{equation}
\theta \, =  \, \frac{u\, f'(u) \, - \, f(u) \, (6\, + \, u \chi'(u))}{2 \sqrt{f(u)}}. 
\end{equation}

Close to the boundary, we can expand the function $h(u)$ as in \eqref{exp} and $f(u) = 1 - M(u) u^3$ (where $M(0)$ would correspond to the energy density of the system) and find that the off-diagonal component of the stress-energy tensor is 
\begin{equation}
T_{xy} \, = \, \frac{1}{2}\left( 3 \, \mathcal{C}_3 \cosh \left( \mathcal{C}_0\right) \, + \, M(0) \sinh (\mathcal{C}_0)\right)\label{stressresult}
\end{equation}
This result is a direct manifestation of the presence of a strain deformation in our background and it will encode the corresponding response, \textit{i.e.} the shear component of the stress.\\
Interestingly one can also notice that 
\begin{equation}
T_x^{ \ y} \, = \, \frac{3}{2} \mathcal{C}_3.
\end{equation}
Using the standard holographic renormalization techniques \cite{Skenderis:2002wp} we can also identify 
\begin{equation}
T_{\mu \nu} \, = \, \frac{3}{2} g_{\mu \nu}^{(3)},
\end{equation}
where $g_{\mu \nu}^{(3)}$ is the sub-leading term of the induced metric expressed in Fefferman-Graham coordinates. As a first step we have to rewrite our ansatz in the FG form using the coordinate transformation
\begin{equation}
\frac{dz^2}{z^2} \, = \, \frac{du^2}{u^2 \, f(u)}
\end{equation}
where $z$ will now be the holographic (FG) coordinate.

Again, using that the asymptotic behaviour of $f(u)$ is $1 - M(u) u^3$ we can find that for small $z$ we have $u= z - \frac{M(z) z^4}{6}$. Now we can already look at our metric and derive the stress-energy tensor. For instance, for the off-diagonal term, we are interested in, we have 
\begin{align}
&g_{x y} (z)  \, = \,  \frac{1}{u(z)^2} \sinh (h(u(z))) = \\
 & = \, \frac{1}{z^2} \left( \cosh (\mathcal{C}_0) + \left( 3 \, \mathcal{C}_3 \cosh \left( \mathcal{C}_0\right) \, + \, M(0) \sinh (\mathcal{C}_0)\right) z^3 \right), 
\end{align}
where higher orders in $z$ have been suppressed.
We can identify $T_{xy}$ easily in this expression and see that the result is the same we found before in \ref{stressresult}. This result give us a robust definition of the non-linear stress in our system which can be indeed identified as:
\begin{equation}
\sigma\,=\,\frac{1}{2}\left( 3 \, \mathcal{C}_3 \cosh \left( \mathcal{C}_0\right) \, + \, M(0) \sinh (\mathcal{C}_0)\right)
\end{equation}

Since we impose the boundary condition $\mathcal{C}_0 = 0$ the stress, we use in all our computations, is simply defined by $\sigma\,=\,\frac{3}{2}\, \mathcal{C}_3$.

\section{Three-phonon interaction terms in Solid EFTs}\label{appendix:perturbation}

To further illustrate the predictive power of the low energy methods, we show here another nontrivial powerful statement that follows from the EFT construction which, as explained above, applies when translations (and possibly scale invariance) are broken spontaneously.

The  nontrivial statement contained in the EFTs \eqref{action} is simply that once the stress-strain relations are known then the full Lagrangian is fixed. Let us illustrate now how this impacts for instance in the determination of the cubic phonon interactions. 
Assuming that the stress-strain relations both for bulk and shear deformations are known, then one can reconstruct the full form of the function $V(X,Z)$ \cite{Alberte:2018doe}, up to an irrelevant additive constant.

Then, the cubic phonon interactions (around the homogeneous, isotropic equilibrium configuration $\phi^I = x^I$) are obtained by expanding our Lagrangian around the background solution, i.e. $\phi^I = x^I + \pi^I$. At third order in $\pi^I$ we obtain 
\begin{equation}
\begin{split}
V(X,Z) \Rightarrow & \,\, \mathcal{C}_1\, (\partial_i \pi^i_L)^3 + \mathcal{C}_2\, (\partial_i \pi^i_L)\, (\dot{\pi}_j)^2 + \mathcal{C}_3 \,(\partial_i \pi^i_L)\,(\partial_j \pi_k)^2 \\
 & + \mathcal{C}_4 \,\dot{\pi}^i\,\dot{\pi}^j\, \partial_i \pi_j + \mathcal{C}_5\, (\partial_i \pi^i_L)\, (\partial_j \pi_k)\,(\partial_k \pi_j)\,,
\end{split}
\end{equation}
where $\pi^I= \pi^I_L+\pi^I_T$. The terms we find are 
\begin{equation}
\mathcal{C}_1 = \frac{1}{6}\left( 8 \,V_{ZZZ}+12 \,V_{XZZ}+6 \,V_{XXZ}+ V_{XXX}\right) + 4\, V_{ZZ}+2\,V_{XZ}+V_Z\,,
\end{equation}
\begin{equation}
\mathcal{C}_2 = -2\,V_{ZZ}- \frac{1}{2}\,V_{XX} - 4 \,V_Z- 2\,V_{XZ}\,,
\end{equation}
\begin{equation}
\mathcal{C}_3 = \frac{1}{2}\,V_{XX}+\,V_{XZ}\,,
\end{equation}
\begin{equation}
\mathcal{C}_4=2\,V_Z\,,
\end{equation}
\begin{equation}
\mathcal{C}_5=-2\,V_{ZZ}-V_Z-V_{XZ}\,.
\end{equation}

In these expressions, the $X,\,Z$- derivatives of $V$ are evaluated on the undeformed configuration. Moreover, by obtaining the relation between $V(X,Z)$ and the stress-strain curve (such as {\it e.g.} Eq.~\eqref{sigma}) one can relate all these  $V(X,Z)$  derivatives to derivatives of the stress strain curve at the origin, $\sigma'(0)$, $\sigma''(0)$, etc, which are measurable quantities. For instance,  the bulk modulus is ${\cal K} = 4  V_{ZZ} + 2  V_{Z} + 4 V_{XZ}+V_{XX}$, $\mathcal{G} = V_X$ and $\epsilon + p =V_X+2  V_{Z}$.  

This illustrates that the realization of symmetries implies nontrivial  relations between distinct low energy  observables. In this example, the strength of the phonon cubic interactions are determined by the shape of the stress-strain curves.

We can write the 5 $\mathcal{C}$'s as a function of these quantities and we would just need two independent new parameters
\begin{equation}
\mathcal{C}_1 = \frac{\mathcal{K}}{2} + \mathcal{N}\,,
\end{equation}
\begin{equation}
\mathcal{C}_2 = \frac{1}{2}\left(3\,\mathcal{G}-\mathcal{K}-3\,(\epsilon + p)\right))\,,
\end{equation}
\begin{equation}
\mathcal{C}_3 = \frac{1}{4}\left(\mathcal{G}+\mathcal{K}-(\epsilon+p)\right)+\mathcal{M}\,,
\end{equation}
\begin{equation}
    \mathcal{C}_4 = \epsilon + p -\mathcal{G}\,,
\end{equation}
\begin{equation}
    \mathcal{C}_5 = \frac{1}{4}\left(\mathcal{G}-\mathcal{K}-(\epsilon+p)\right) + \mathcal{M}\,.
\end{equation}
where the independent new parameters are 
\begin{equation}
    \mathcal{N} =  \frac{1}{6}\left( 8 \,V_{ZZZ}+12 \,V_{XZZ}+6 \,V_{XXZ}+ V_{XXX}\right) - 2 \, V_{ZZ} - \frac{1}{2}\,V_{XX}\,,
\end{equation}
\begin{equation}
    \mathcal{M} = \frac{1}{4}\,V_{XX} - V_{ZZ}\,.
\end{equation}
We can compare our results with Ref. \cite{Leutwyler:1996er}. There he concludes that there are 3 independent new parameters, but we think the difference comes from the fact that he is working in 3 space-dimensions instead of 2. Moreover, in Ref. \cite{Leutwyler:1996er} there are 6 independent operators in the cubic expansion. It is trivial to check that in two dimensions the extra operator can be expressed as a function of the others
\begin{equation}
(\partial_i \pi_j)\,(\partial_i\pi_k)\,(\partial_j\pi_k) = (\partial_i\pi^i)\,(\partial_j\pi_k)^2 + \frac{(\partial_i\pi^i)}{2}\,\left((\partial_j\pi_k)(\partial_k\pi_j)-(\partial_i\pi^i)^2\right)\,.
\end{equation}

In the case of scale invariance these terms simplify considerably. Let us then take $V(X,Z) = Z^{\frac{1}{2}+\omega}\,f\left(\frac{X}{\sqrt{Z}}\right)$
\begin{equation}
\mathcal{C}_1 = \frac{\omega}{3}\left((2+6\,\omega + 4 \,\omega^2) f(1) - 3 f'(1)\right)\,,
\end{equation}
\begin{equation}
\mathcal{C}_2 = \frac{1}{2}\left(-(3+8\,\omega+4\,\omega^2) \,f(1)+3\,f'(1)\right)\,,
\end{equation}
\begin{equation}
\mathcal{C}_3 = \omega\,f'(1)\,,
\end{equation}
\begin{equation}
\mathcal{C}_4=(1+2\omega)\,f(1)-f'(1)\,,
\end{equation}
\begin{equation}
\mathcal{C}_5=\omega\,(f'(1)-(1+2\,\omega)\,f(1))\,.
\end{equation}

This implies that for a scale invariant potential there are no free parameters: we can identify $f(1)$, $f'(1)$ and $\omega$ with $\mathcal{K}$, $\mathcal{G}$ and $\epsilon + p$
 \begin{equation}
     \mathcal{G} = f'(1)\,,
 \end{equation}
 \begin{equation}
     \mathcal{K} = 2\,\omega\,(1+2\,\omega)\,f(1)\,,
 \end{equation}
 \begin{equation}
     \epsilon + p = (1+2\,\omega)\,f(1)\,.
 \end{equation}
 Therefore cubic interactions are all fixed by these background or linear elasticity quantities. This is true both for general SI as well as conformal solid limit (which is just a particular value of $w$).
\bibliographystyle{JHEP}
\bibliography{NLE}

\providecommand{\href}[2]{#2}\begingroup\raggedright\begin{thebibliography}{10}

\bibitem{landau7}
L.~D. Landau and E.~M. Lifshitz, \emph{Course of Theoretical Physics, Vol.
  7,Theory of Elasticity}.
\newblock Pergamon Press, 1970.

\bibitem{Lubensky}
P.~M. Chaikin and T.~C. Lubensky, \emph{Principles of Condensed Matter
  Physics}.
\newblock Cambridge University Press, 1995,
  \href{http://dx.doi.org/10.1017/CBO9780511813467}{10.1017/CBO9780511813467}.

\bibitem{Ogden2004}
R.~W. Ogden, G.~Saccomandi and I.~Sgura, \emph{Fitting hyperelastic models to
  experimental data},
  \href{http://dx.doi.org/10.1007/s00466-004-0593-y}{\emph{Computational
  Mechanics} {\bf 34} (Nov, 2004) 484--502}.

\bibitem{ZAMM:ZAMM19850650903}
R.~W. Ogden, \emph{Non-Linear Elastic Deformations}.
\newblock WILEY-VCH Verlag, 1985.

\bibitem{PhysRevA.6.2401}
P.~C. Martin, O.~Parodi and P.~S. Pershan, \emph{Unified hydrodynamic theory
  for crystals, liquid crystals, and normal fluids},
  \href{http://dx.doi.org/10.1103/PhysRevA.6.2401}{\emph{Phys. Rev. A} {\bf 6}
  (Dec, 1972) 2401--2420}.

\bibitem{enlighten70333}
Y.B.Fu and R.~Ogden, \emph{Nonlinear Elasticity: Theory and Applications}.
\newblock No.~283 in London Mathematical Society lecture note series. Cambridge
  University Press, Cambridge, UK, 2001,
  \href{http://dx.doi.org/10.1017/CBO9780511526466}{10.1017/CBO9780511526466}.

\bibitem{Leutwyler:1996er}
H.~Leutwyler, \emph{{Phonons as goldstone bosons}}, {\emph{Helv. Phys. Acta}
  {\bf 70} (1997) 275--286}, [\href{https://arxiv.org/abs/hep-ph/9609466}{{\tt
  hep-ph/9609466}}].

\bibitem{Dubovsky:2011sj}
S.~Dubovsky, L.~Hui, A.~Nicolis and D.~T. Son, \emph{{Effective field theory
  for hydrodynamics: thermodynamics, and the derivative expansion}},
  \href{http://dx.doi.org/10.1103/PhysRevD.85.085029}{\emph{Phys. Rev.} {\bf
  D85} (2012) 085029}, [\href{https://arxiv.org/abs/1107.0731}{{\tt
  1107.0731}}].

\bibitem{Nicolis:2013lma}
A.~Nicolis, R.~Penco and R.~A. Rosen, \emph{{Relativistic Fluids, Superfluids,
  Solids and Supersolids from a Coset Construction}},
  \href{http://dx.doi.org/10.1103/PhysRevD.89.045002}{\emph{Phys. Rev.} {\bf
  D89} (2014) 045002}, [\href{https://arxiv.org/abs/1307.0517}{{\tt
  1307.0517}}].

\bibitem{Nicolis:2015sra}
A.~Nicolis, R.~Penco, F.~Piazza and R.~Rattazzi, \emph{{Zoology of condensed
  matter: Framids, ordinary stuff, extra-ordinary stuff}},
  \href{http://dx.doi.org/10.1007/JHEP06(2015)155}{\emph{JHEP} {\bf 06} (2015)
  155}, [\href{https://arxiv.org/abs/1501.03845}{{\tt 1501.03845}}].

\bibitem{Alberte:2018doe}
L.~Alberte, M.~Baggioli, V.~Cancer-Castillo and O.~Pujolas, \emph{{Elasticity
  bounds from Effective Field Theory}},
  \href{http://dx.doi.org/10.1103/PhysRevD.100.065015}{\emph{Phys. Rev.} {\bf
  D100} (2019) 065015}, [\href{https://arxiv.org/abs/1807.07474}{{\tt
  1807.07474}}].

\bibitem{Baggioli:2019elg}
M.~Baggioli, V.~C. Castillo and O.~Pujolas, \emph{{Scale invariant solids}},
  \href{https://arxiv.org/abs/1910.05281}{{\tt 1910.05281}}.

\bibitem{Boyle:2018uiv}
L.~Boyle, M.~Dickens and F.~Flicker, \emph{{Conformal Quasicrystals and
  Holography}},
  \href{http://dx.doi.org/10.1103/PhysRevX.10.011009}{\emph{Phys.\ Rev.\ X}
  {\bf 10} (2020) 011009}, [\href{https://arxiv.org/abs/1805.02665}{{\tt
  1805.02665}}].

\bibitem{PhysRevResearch.2.022022}
M.~Baggioli, \emph{Homogeneous holographic viscoelastic models and
  quasicrystals},
  \href{http://dx.doi.org/10.1103/PhysRevResearch.2.022022}{\emph{Phys. Rev.
  Research} {\bf 2} (Apr, 2020) 022022}.

\bibitem{Baggioli:2019rrs}
M.~Baggioli, \emph{{Applied Holography}: {A Practical Mini-Course}}.
\newblock SpringerBriefs in Physics. Springer, 2019,
  \href{http://dx.doi.org/10.1007/978-3-030-35184-7}{10.1007/978-3-030-35184-7}.

\bibitem{Emparan:2009at}
R.~Emparan, T.~Harmark, V.~Niarchos and N.~A. Obers, \emph{{Essentials of
  Blackfold Dynamics}},
  \href{http://dx.doi.org/10.1007/JHEP03(2010)063}{\emph{JHEP} {\bf 03} (2010)
  063}, [\href{https://arxiv.org/abs/0910.1601}{{\tt 0910.1601}}].

\bibitem{Emparan:2016sjk}
R.~Emparan, K.~Izumi, R.~Luna, R.~Suzuki and K.~Tanabe, \emph{{Hydro-elastic
  Complementarity in Black Branes at large D}},
  \href{http://dx.doi.org/10.1007/JHEP06(2016)117}{\emph{JHEP} {\bf 06} (2016)
  117}, [\href{https://arxiv.org/abs/1602.05752}{{\tt 1602.05752}}].

\bibitem{Vegh:2013sk}
D.~Vegh, \emph{{Holography without translational symmetry}},
  \href{https://arxiv.org/abs/1301.0537}{{\tt 1301.0537}}.

\bibitem{Blake:2013owa}
M.~Blake, D.~Tong and D.~Vegh, \emph{{Holographic Lattices Give the Graviton an
  Effective Mass}},
  \href{http://dx.doi.org/10.1103/PhysRevLett.112.071602}{\emph{Phys. Rev.
  Lett.} {\bf 112} (2014) 071602}, [\href{https://arxiv.org/abs/1310.3832}{{\tt
  1310.3832}}].

\bibitem{Andrade:2013gsa}
T.~Andrade and B.~Withers, \emph{{A simple holographic model of momentum
  relaxation}}, \href{http://dx.doi.org/10.1007/JHEP05(2014)101}{\emph{JHEP}
  {\bf 05} (2014) 101}, [\href{https://arxiv.org/abs/1311.5157}{{\tt
  1311.5157}}].

\bibitem{Baggioli:2014roa}
M.~Baggioli and O.~Pujolas, \emph{{Electron-Phonon Interactions,
  Metal-Insulator Transitions, and Holographic Massive Gravity}},
  \href{http://dx.doi.org/10.1103/PhysRevLett.114.251602}{\emph{Phys. Rev.
  Lett.} {\bf 114} (2015) 251602}, [\href{https://arxiv.org/abs/1411.1003}{{\tt
  1411.1003}}].

\bibitem{Alberte:2015isw}
L.~Alberte, M.~Baggioli, A.~Khmelnitsky and O.~Pujolas, \emph{{Solid Holography
  and Massive Gravity}},
  \href{http://dx.doi.org/10.1007/JHEP02(2016)114}{\emph{JHEP} {\bf 02} (2016)
  114}, [\href{https://arxiv.org/abs/1510.09089}{{\tt 1510.09089}}].

\bibitem{Ammon:2020xyv}
M.~Ammon, M.~Baggioli, S.~Gray, S.~Grieninger and A.~Jain, \emph{{On the
  Hydrodynamic Description of Holographic Viscoelastic Models}},
  \href{https://arxiv.org/abs/2001.05737}{{\tt 2001.05737}}.

\bibitem{Alberte:2017cch}
L.~Alberte, M.~Ammon, M.~Baggioli, A.~Jiménez and O.~Pujolàs, \emph{{Black
  hole elasticity and gapped transverse phonons in holography}},
  \href{http://dx.doi.org/10.1007/JHEP01(2018)129}{\emph{JHEP} {\bf 01} (2018)
  129}, [\href{https://arxiv.org/abs/1708.08477}{{\tt 1708.08477}}].

\bibitem{Alberte:2017oqx}
L.~Alberte, M.~Ammon, M.~Baggioli, A.~Jiménez-Alba and O.~Pujolàs,
  \emph{{Holographic Phonons}},  \href{https://arxiv.org/abs/1711.03100}{{\tt
  1711.03100}}.

\bibitem{Alberte:2016xja}
L.~Alberte, M.~Baggioli and O.~Pujolas, \emph{{Viscosity bound violation in
  holographic solids and the viscoelastic response}},
  \href{http://dx.doi.org/10.1007/JHEP07(2016)074}{\emph{JHEP} {\bf 07} (2016)
  074}, [\href{https://arxiv.org/abs/1601.03384}{{\tt 1601.03384}}].

\bibitem{Baggioli:2018bfa}
M.~Baggioli and A.~Buchel, \emph{{Holographic Viscoelastic Hydrodynamics}},
  \href{https://arxiv.org/abs/1805.06756}{{\tt 1805.06756}}.

\bibitem{Andrade:2019zey}
T.~Andrade, M.~Baggioli and O.~Pujolas, \emph{{Viscoelastic Dynamics in
  Holography}},  \href{https://arxiv.org/abs/1903.02859}{{\tt 1903.02859}}.

\bibitem{Baggioli:2019mck}
M.~Baggioli, S.~Grieninger and H.~Soltanpanahi, \emph{{Nonlinear Oscillatory
  Shear Tests in Viscoelastic Holography}},
  \href{https://arxiv.org/abs/1910.06331}{{\tt 1910.06331}}.

\bibitem{Baggioli:2020nay}
M.~Baggioli, \emph{{Are The Homogeneous Holographic Viscoelastic Models
  Quasicrystals ?}},  \href{https://arxiv.org/abs/2001.06228}{{\tt
  2001.06228}}.

\bibitem{Baggioli:2016rdj}
M.~Baggioli, \emph{{Gravity, holography and applications to condensed matter}}.
\newblock PhD thesis, Barcelona U., 2016.
\newblock \href{https://arxiv.org/abs/1610.02681}{{\tt 1610.02681}}.

\bibitem{Baggioli:2015zoa}
M.~Baggioli and M.~Goykhman, \emph{{Phases of holographic superconductors with
  broken translational symmetry}},
  \href{http://dx.doi.org/10.1007/JHEP07(2015)035}{\emph{JHEP} {\bf 07} (2015)
  035}, [\href{https://arxiv.org/abs/1504.05561}{{\tt 1504.05561}}].

\bibitem{Baggioli:2015gsa}
M.~Baggioli and D.~K. Brattan, \emph{{Drag phenomena from holographic massive
  gravity}},
  \href{http://dx.doi.org/10.1088/1361-6382/34/1/015008}{\emph{Class. Quant.
  Grav.} {\bf 34} (2017) 015008}, [\href{https://arxiv.org/abs/1504.07635}{{\tt
  1504.07635}}].

\bibitem{Baggioli:2015dwa}
M.~Baggioli and M.~Goykhman, \emph{{Under The Dome: Doped holographic
  superconductors with broken translational symmetry}},
  \href{http://dx.doi.org/10.1007/JHEP01(2016)011}{\emph{JHEP} {\bf 01} (2016)
  011}, [\href{https://arxiv.org/abs/1510.06363}{{\tt 1510.06363}}].

\bibitem{Baggioli:2020ljz}
M.~Baggioli and W.-J. Li, \emph{{Universal Bounds on Transport in Holographic
  Systems with Broken Translations}},
  \href{https://arxiv.org/abs/2005.06482}{{\tt 2005.06482}}.

\bibitem{Bardoux:2012aw}
Y.~Bardoux, M.~M. Caldarelli and C.~Charmousis, \emph{{Shaping black holes with
  free fields}}, \href{http://dx.doi.org/10.1007/JHEP05(2012)054}{\emph{JHEP}
  {\bf 05} (2012) 054}, [\href{https://arxiv.org/abs/1202.4458}{{\tt
  1202.4458}}].

\bibitem{Caldarelli:2016nni}
M.~M. Caldarelli, A.~Christodoulou, I.~Papadimitriou and K.~Skenderis,
  \emph{{Phases of planar AdS black holes with axionic charge}},
  \href{http://dx.doi.org/10.1007/JHEP04(2017)001}{\emph{JHEP} {\bf 04} (2017)
  001}, [\href{https://arxiv.org/abs/1612.07214}{{\tt 1612.07214}}].

\bibitem{Ammon:2019apj}
M.~Ammon, M.~Baggioli, S.~Gray and S.~Grieninger, \emph{{Longitudinal Sound and
  Diffusion in Holographic Massive Gravity}},
  \href{http://dx.doi.org/10.1007/JHEP10(2019)064}{\emph{JHEP} {\bf 10} (2019)
  064}, [\href{https://arxiv.org/abs/1905.09164}{{\tt 1905.09164}}].

\bibitem{Baggioli:2019abx}
M.~Baggioli and S.~Grieninger, \emph{{Zoology of solid \& fluid holography —
  Goldstone modes and phase relaxation}},
  \href{http://dx.doi.org/10.1007/JHEP10(2019)235}{\emph{JHEP} {\bf 10} (2019)
  235}, [\href{https://arxiv.org/abs/1905.09488}{{\tt 1905.09488}}].

\bibitem{Kachru:2008yh}
S.~Kachru, X.~Liu and M.~Mulligan, \emph{{Gravity duals of Lifshitz-like fixed
  points}}, \href{http://dx.doi.org/10.1103/PhysRevD.78.106005}{\emph{Phys.
  Rev. D} {\bf 78} (2008) 106005}, [\href{https://arxiv.org/abs/0808.1725}{{\tt
  0808.1725}}].

\bibitem{Taylor:2008tg}
M.~Taylor, \emph{{Non-relativistic holography}},
  \href{https://arxiv.org/abs/0812.0530}{{\tt 0812.0530}}.

\bibitem{Cremonini:2014pca}
S.~Cremonini, X.~Dong, J.~Rong and K.~Sun, \emph{{Holographic RG flows with
  nematic IR phases}},
  \href{http://dx.doi.org/10.1007/JHEP07(2015)082}{\emph{JHEP} {\bf 07} (2015)
  082}, [\href{https://arxiv.org/abs/1412.8638}{{\tt 1412.8638}}].

\bibitem{Bhattacharya:2014dea}
J.~Bhattacharya, S.~Cremonini and B.~Goutéraux, \emph{{Intermediate scalings
  in holographic RG flows and conductivities}},
  \href{http://dx.doi.org/10.1007/JHEP02(2015)035}{\emph{JHEP} {\bf 02} (2015)
  035}, [\href{https://arxiv.org/abs/1409.4797}{{\tt 1409.4797}}].

\bibitem{Mateos:2011ix}
D.~Mateos and D.~Trancanelli, \emph{{The anisotropic N=4 super Yang-Mills
  plasma and its instabilities}},
  \href{http://dx.doi.org/10.1103/PhysRevLett.107.101601}{\emph{Phys. Rev.
  Lett.} {\bf 107} (2011) 101601}, [\href{https://arxiv.org/abs/1105.3472}{{\tt
  1105.3472}}].

\bibitem{Jain:2014vka}
S.~Jain, N.~Kundu, K.~Sen, A.~Sinha and S.~P. Trivedi, \emph{{A Strongly
  Coupled Anisotropic Fluid From Dilaton Driven Holography}},
  \href{http://dx.doi.org/10.1007/JHEP01(2015)005}{\emph{JHEP} {\bf 01} (2015)
  005}, [\href{https://arxiv.org/abs/1406.4874}{{\tt 1406.4874}}].

\bibitem{Rebhan:2011vd}
A.~Rebhan and D.~Steineder, \emph{{Violation of the Holographic Viscosity Bound
  in a Strongly Coupled Anisotropic Plasma}},
  \href{http://dx.doi.org/10.1103/PhysRevLett.108.021601}{\emph{Phys. Rev.
  Lett.} {\bf 108} (2012) 021601}, [\href{https://arxiv.org/abs/1110.6825}{{\tt
  1110.6825}}].

\bibitem{Baggioli:2016pia}
M.~Baggioli, B.~Goutéraux, E.~Kiritsis and W.-J. Li, \emph{{Higher derivative
  corrections to incoherent metallic transport in holography}},
  \href{http://dx.doi.org/10.1007/JHEP03(2017)170}{\emph{JHEP} {\bf 03} (2017)
  170}, [\href{https://arxiv.org/abs/1612.05500}{{\tt 1612.05500}}].

\bibitem{Baggioli:2017ojd}
M.~Baggioli and W.-J. Li, \emph{{Diffusivities bounds and chaos in holographic
  Horndeski theories}},
  \href{http://dx.doi.org/10.1007/JHEP07(2017)055}{\emph{JHEP} {\bf 07} (2017)
  055}, [\href{https://arxiv.org/abs/1705.01766}{{\tt 1705.01766}}].

\bibitem{Biasi:2019eap}
A.~Biasi, J.~Mas and A.~Serantes, \emph{{Gravitational wave driving of a gapped
  holographic system}},
  \href{http://dx.doi.org/10.1007/JHEP05(2019)161}{\emph{JHEP} {\bf 05} (2019)
  161}, [\href{https://arxiv.org/abs/1903.05618}{{\tt 1903.05618}}].

\bibitem{ishii2019glass}
Y.~Ishii, Y.~Ouchi, S.~Kawaguchi, H.~Ishibashi, Y.~Kubota and S.~Mori,
  \emph{Glass-like features of crystalline solids in the quantum critical
  regime}, {\emph{arXiv preprint arXiv:1901.09502} (2019) }.

\bibitem{setty2019glass}
C.~Setty, \emph{Glass-induced enhancement of superconducting $ t\_c $: Pairing
  via dissipative mediators}, {\emph{arXiv preprint arXiv:1902.00516} (2019) }.

\bibitem{Baggioli:2018qwu}
M.~Baggioli and A.~Zaccone, \emph{{Universal origin of boson peak vibrational
  anomalies in ordered crystals and in amorphous materials}},
  \href{https://arxiv.org/abs/1810.09516}{{\tt 1810.09516}}.

\bibitem{baggioli2018soft}
M.~Baggioli and A.~Zaccone, \emph{Soft optical phonons induce glassy-like
  vibrational and thermal anomalies in ordered crystals}, {\emph{arXiv preprint
  arXiv:1812.07245} (2018) }.

\bibitem{baggioli2019unified}
M.~Baggioli and A.~Zaccone, \emph{Unified theory of vibrational spectra in
  amorphous materials},  2019.

\bibitem{Scheuerlein:2017kjv}
C.~Scheuerlein, F.~Lackner, F.~Savary, B.~Rehmer, M.~Finn and C.~Meyer,
  \emph{{Thermomechanical behavior of the HL-LHC 11 Tesla Nb$_{3}$Sn magnet
  coil constituents during reaction heat treatment}},
  \href{https://arxiv.org/abs/1711.07022}{{\tt 1711.07022}}.

\bibitem{Scheuerlein:2017zka}
C.~Scheuerlein, F.~Lackner, F.~Savary, B.~Rehmer, M.~Finn and P.~Uhlemann,
  \emph{{Mechanical Properties of the HL-LHC 11 T Nb$_3$Sn Magnet Constituent
  Materials}}, \href{http://dx.doi.org/10.1109/TASC.2016.2638046}{\emph{IEEE
  Trans. Appl. Supercond.} {\bf 27} (2017) 4003007}.

\bibitem{Balasubramanian:1999re}
V.~Balasubramanian and P.~Kraus, \emph{{A Stress tensor for Anti-de Sitter
  gravity}}, \href{http://dx.doi.org/10.1007/s002200050764}{\emph{Commun. Math.
  Phys.} {\bf 208} (1999) 413--428},
  [\href{https://arxiv.org/abs/hep-th/9902121}{{\tt hep-th/9902121}}].

\bibitem{Skenderis:2002wp}
K.~Skenderis, \emph{{Lecture notes on holographic renormalization}},
  \href{http://dx.doi.org/10.1088/0264-9381/19/22/306}{\emph{Class. Quant.
  Grav.} {\bf 19} (2002) 5849--5876},
  [\href{https://arxiv.org/abs/hep-th/0209067}{{\tt hep-th/0209067}}].

\end{thebibliography}\endgroup
\end{document}